\pdfoutput=1
\PassOptionsToPackage{prologue,table}{xcolor} 
\documentclass[sigconf,screen,nonacm]{acmart}

\usepackage{xcolor}
\usepackage{graphicx}
\usepackage{subcaption}
\usepackage[capitalise]{cleveref}
\usepackage{pgfplots}
\usepackage{pgfplotstable}
\usepackage{tikz}
\usepackage{fontawesome5}
\usepackage{algpseudocode}
\usepackage{algorithm}
\usepackage{microtype}
\usepackage{enumerate}

\usetikzlibrary{patterns,positioning} 
\pgfplotsset{compat=1.18}

\tikzset{
    node/.style={circle, draw, minimum size=8mm, inner sep=0},
    rednode/.style={circle, draw, minimum size=8mm, inner sep=0, fill=red!20},
    edge/.style={->, thick},
    weight/.style={font=\small, midway, above}
}

\newcommand{\numbercircle}[1]{%
  \tikz[baseline=(char.base)]{
    \node[shape=circle,draw,inner sep=0.5pt,fill=black,text=white,scale=1] (char) {#1};
  }%
}

\setcopyright{acmlicensed}
\copyrightyear{2025}
\acmYear{2025}
\acmDOI{XXXXXXX.XXXXXXX}

\acmConference[HPDC '25]{Make sure to enter the correct
  conference title from your rights confirmation email}{July 20--23,
  2025}{Notre Dame, IN, USA}
\acmBooktitle{The 34th International Symposium on
High-Performance Parallel and Distributed Computing (HPDC '25), July 20--23, 2025, Notre Dame, IN, USA}
\acmISBN{978-1-4503-XXXX-X/18/06}

\begin{document}


\title{\textsc{Databelt}: A Continuous Data Path for Serverless Workflows in the 3D~Compute Continuum}


\author{Cynthia Marcelino}
\orcid{0000-0003-1707-3014}
\affiliation{%
  \department{Distributed Systems Group}
  \city{TU Wien, Vienna}
  \country{Austria}
}

\author{Leonard Guelmino}
\orcid{0009-0007-8686-5492}
\affiliation{%
  \department{Distributed Systems Group}
  \city{TU Wien, Vienna}
  \country{Austria}
  }

\author{Thomas Pusztai}
\orcid{0000-0001-9765-6310}
\affiliation{%
  \department{Distributed Systems Group}
  \city{TU Wien, Vienna}
  \country{Austria}
  }

\author{Stefan Nastic}
\orcid{0000-0003-0410-6315}
\affiliation{%
    \department{Distributed Systems Group}
    \city{TU Wien, Vienna}
    \country{Austria}
  }


\begin{abstract}
  Serverless computing allows for dynamic and flexible execution of FaaS functions while simplifying infrastructure management. Typically, serverless functions rely on remote storage services for managing state, which can result in increased latency and network communication overhead. In a dynamic environment such as the 3D (Edge-Cloud-Space) Compute Continuum, serverless functions face additional challenges due to frequent changes in network topology. As satellites move in and out of the range of ground stations, functions must make multiple hops to access cloud services, leading to high-latency state access and unnecessary data transfers.

  In this paper, we present Databelt, a state management framework for serverless workflows designed for the dynamic environment of the 3D Compute Continuum. Databelt introduces an SLO-aware state propagation mechanism that enables the function state to move continuously in orbit. Databelt proactively offloads function states to the most suitable node, such that when functions execute, the data is already present on the execution node or nearby, thus minimizing state access latency and reducing the number of network hops. Additionally, Databelt introduces a function state fusion mechanism that abstracts state management for functions sharing the same serverless runtime. When functions are fused, Databelt seamlessly retrieves their state as a group, reducing redundant network and storage operations and improving overall workflow efficiency.
      
  Our experimental results show that Databelt reduces workflow execution time by up to 66\% and increases throughput by 50\% compared to the baselines. 
  Furthermore, our results show that Databelt function state fusion reduces storage operations latency by up to 20\%, by reducing repetitive storage requests for functions within the same runtime, ensuring efficient execution of serverless workflows in highly dynamic network environments such as the 3D Continuum.
\end{abstract}


\maketitle


\section{Introduction} \label{sec:intro}

Over the past few years, we have seen a massive increase in the number of low Earth orbit (LEO) satellites, with more than 8,000 total in orbit in 2024~\cite{LEO_SAT_stats}.
Those numbers are going to increase further as, e.g., Starlink intends to grow beyond 30,000~satellites~\cite{Starlink30K_SatsPaperwork2019} and Amazon has begun launching its Kuiper constellation with more than 3,000~planned satellites~\cite{FCC_KuiperAuthorization2020}.
The satellites of such megaconstellations are interconnected with inter-satellite laser links (ISLs)~\cite{NetworkTopology2020}, thus establishing a high-speed and high-bandwidth communication network in space.
Nowadays, most LEO satellites are employed for communication purposes using a ``bent-pipe'' architecture, where a signal from a transmitter on Earth is relayed directly or through some ISLs (``extended bent-pipe'') to a receiver~\cite{StarlinkPerf2024}. However, ``bent-pipes'' face scalability limitations as megaconstellations and data volumes increase~\cite{OEC_ML}.

As the computing capacity of new satellite generations increases, these megaconstellations are projected to become an attractive source of low-latency computational power that is available anywhere on Earth~\cite{SatelliteComputingVision2023, Vision6G2024}.
The visions of LEO Edge computing~\cite{MobileEdgeCompLEO_AntColony2022,LeoComputingPlatform2021,JointHugeLEO,DynamicTransmissionLEOMEC}, Satellite Computing~\cite{SCCaseStudy2023, SatelliteComputingVision2023,EnergySatDesign}, or Orbital Edge Computing (OEC)~\cite{OECNano2020,OECSurvey2023,ComputationalOEC} share the idea that LEO satellites offer useful computing capacities in space for a variety of use cases. However, they consider LEO as a separate layer, focusing primarily on offloading tasks to satellites~\cite{DNN2023,optimizingFLscheduling2023,OrbitaEdgeOffloading2022}. 
The 3D~Continuum~\cite{HyperDrive2024, Cosmos2025} introduces a unified Edge-Cloud-Space computing continuum that allows running workloads seamlessly across the entire range of nodes. 

Serverless computing is a suitable paradigm for the 3D~Continuum, but due to its stateless design, data management becomes challenging~\cite{HyperDrive2024, Cosmos2025, LeoComputingPlatform2021}.
In serverless computing, functions typically store their state data in remote storage services, and the state data must be transferred to the functions that require it, potentially incurring a large impact on the performance of the workflow~\cite{Cloudburst2020,sonic, Boki,faasm}.
Moreover, every function typically fulfills one task and can be seen as a reusable building block for complex applications, which are designed by composing multiple functions in a workflow~\cite{goldfish2024,CloudProgrammingSimplified}.

\begin{figure}[t]
    \centering
    \includegraphics[width=0.99\linewidth]{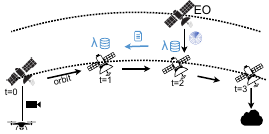}
    \caption{Function states propagation to counter orbital movements to minimize network hops, while data from the EO satellite is fetched.}
    \label{fig:motivation}
\end{figure}

Due to their high-velocity orbits, satellites are in direct range of a ground station or another satellite for a short time only (\cref{fig:motivation}).
Therefore, the realization of stateful serverless workflows becomes complex because state management must consider dynamically changing latencies.
Although ISL communication is fast and has low latency, placing data on a poorly selected satellite could mean that a function must retrieve its state from the other side of the Earth, which incurs unnecessarily high latency.

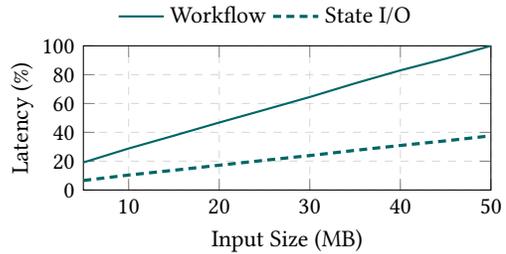
\begin{figure}[t]  
\centering
    \begin{tikzpicture}
        \begin{axis}[               
            xlabel={Input Size (MB)},
            ylabel style={yshift=-3pt},
            ylabel={Latency (\%)},
            xtick={10,20,30,40,50}, 
            xmin=5,
            xmax=50,
            ymin=0,
            ymax=100,
            legend style={at={(0.45,1.35)},anchor=north,draw=none,legend columns=-1},
            width=7cm,         
            height=3.5cm, 
            grid=major,
            grid style={dashed,gray!30},
            mark options={solid}
        ]
        \addplot[
            color=teal!80!black,
            thick,
        ] coordinates {
            (5, 19.1) (10, 28.8) (15, 37.7) (20, 46.8) (25, 55.6) (30, 64.5) (35, 74.0) (40, 83.0) (45, 91.1) (50, 100.0)
        };
        \addlegendentry{Workflow}

        \addplot[
            color=teal!80!black,
            very thick,
            dash pattern=on 3pt off 2pt,
        ] coordinates {
            (5, 6.6) (10, 10.4) (15, 13.7) (20, 17.2) (25, 20.6) (30, 23.9) (35, 27.5) (40, 30.9) (45, 34.1) (50, 37.6)
        };
        \addlegendentry{State I/O}
        \end{axis}
    \end{tikzpicture}
    \caption{Normalized workflow and state management latencies in a serverless workflow, showing the contribution of state I/O~(dashed) to total workflow latency~(solid) across varying input sizes.}
    \label{fig:motivation_cdf}
\end{figure}

To investigate the time required to transfer state to/from functions, we design an experiment that records the reading and writing of function state and the total workflow execution time. Our workflow consists of four functions and relies on an external key-value store (KVS) to store the functions' states (detailed experiment description is presented in \cref{sec:evaluation}).
\cref{fig:motivation_cdf} shows the normalized latencies of the overall workflow and the state I/O.
The graph shows that I/O contributes up to 40\% of the total workflow latency.
Research reports function interaction overheads (which include but are not limited to I/O) reaching up to 95\% of the execution time of a single function~\cite{SAND2018, Faastlane2021, Truffle2024}.
Naturally, the issue with I/O overhead is exacerbated in the 3D~Continuum where nodes have dynamic positions. Therefore, a serverless platform for the 3D~Continuum must consider when satellites are going to be in and out of range and place or prefetch data accordingly.

Stateful serverless systems for the Cloud~\cite{DurableFunctions2021, Cloudburst2020, Crucial2022, AWS_StepFunctionsOverview,faasTrack} typically assume a low-latency connection to the remote storage services, which is usually not the case in the 3D~Continuum.
Serverless systems for the Edge, e.g.,~\cite{TAROT,FaasModelForEdge2022, goldfish2024, DS2P2024}, are often designed to reduce the latency for state access by elastically scheduling functions on nodes that are close to the data, keeping some of the data on the compute nodes, and/or migrating data.
However, they typically fail to account for the fast-paced network topology changes that are inherent to the 3D~Continuum, because LEO satellites move at about 27,000~km/h.
State placement and prefetching must account for when a satellite will come into range or go out of range.
Serverless platforms for LEO are just starting to surface, and to the best of our knowledge, most of them ~\cite{Krios,HyperDrive2024, komet2024, ResilientAccessEquality6G_LEO2023} either do not consider generic function state data or handle only single functions instead of workflows, or do not specify how they handle data.

ISLs allow satellites to communicate without requiring direct line-of-sight to a ground station or another satellite, decreasing latency and alleviating network congestion. However, to minimize network hops and optimize end-to-end latency, the function and its state must be strategically placed within the accessible range of other functions in the workflow.
We summarize our contributions as follows:
\begin{itemize}

    \item \textit{Databelt: A novel state-aware serverless model and architecture} that enables state placement in the dynamic and heterogeneous environments of the 3D Continuum. Databelt allows serverless functions to move the data in orbit closer to the target function, and place functions within a workflow on nearby nodes,  reducing workflow latency while adhering to specific environmental conditions of the 3D Compute Continuum;

    \item \textit{A function state propagation mechanism} that leverages node position, including edge, cloud, and satellite, to identify the neighbor nodes and propagate the state to specific nodes within the execution range while complying with SLO requirements. Hence, providing local function state availability by up to 79\%, and decreasing the workflow latency by up to 66\% compared to random storage placement strategies;
     
    \item  \textit{A function state fusion mechanism} to avoid multiple state retrievals for functions that share the same serverless runtime, thus minimizing storage operations to a constant request amount instead of linear increase and consequently reducing latency by up to 20\% compared to multiple single state access retrievals.
\end{itemize}

This paper has eight sections. 
\cref{sec:motivation} presents an illustrative scenario and background.
\cref{sec:architecture} presents formalized key requirements of the 3D Continuum, and an overview of the Databelt's architecture. 
\cref{sec:mechanisms} describes the mechanisms introduced by Databelt. 
\cref{sec:impl} describes prototype implementation.
\cref{sec:evaluation} presents experiments, and evaluation, 
\cref{sec:relatedw} presents related work. 
\cref{sec:conclusion} concludes with a final discussion and future work.


\section{Motivation \& Background} \label{sec:motivation}

In this section, we present a realistic use case and background information on our previous work on scheduling that Databelt builds upon.

\subsection{Illustrative Scenario}\label{subsec:scenario}

\begin{figure}[t]
    \centering
    \includegraphics[width=\linewidth]{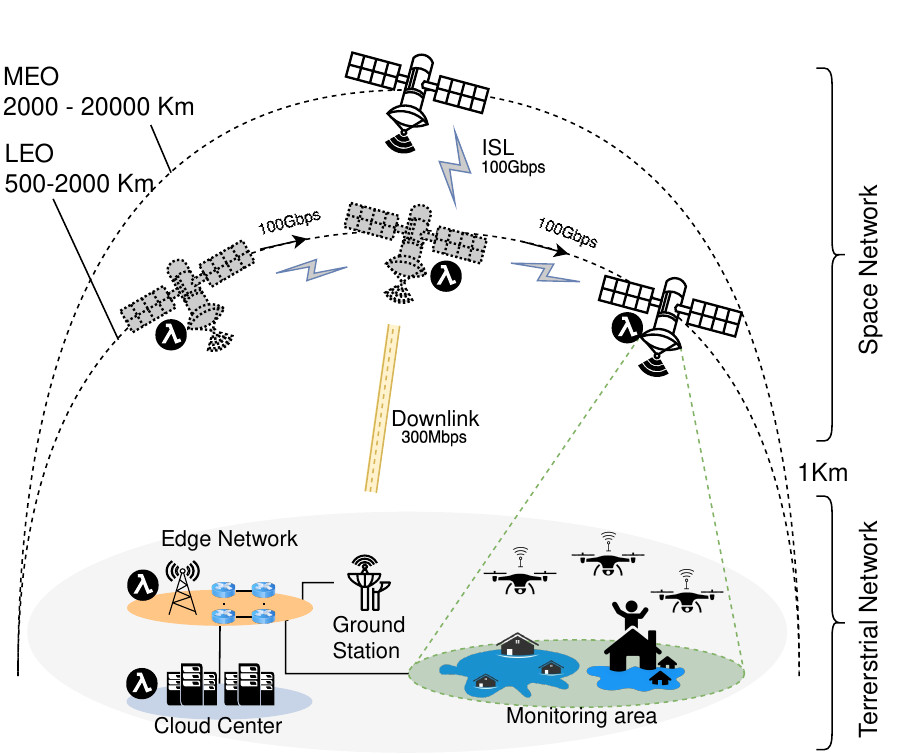}
    \caption{Flood Disaster Detection Illustrative Scenario.}
    \label{fig:use-case}
\end{figure}

To further motivate the need for stateful serverless computing in the 3D Continuum we describe an illustrative scenario for flood disaster detection.
\cref{fig:use-case} shows a realistic use case to support mapping the effects of a flood disaster and finding people in need of rescue.
Autonomous drones fly over a village or small city that is affected by a flood to locate people in distress.
Such areas often lack reliable 5G connectivity, because the natural disaster may have cut base stations off the power grid or otherwise damaged them.
Thus, the drones send their recorded video to a cluster of LEO satellites for rapid detection of people.
The LEO satellites also receive data from an Earth Observation (EO) satellite surveying the affected area with  Synthetic Aperture Radar
(SAR), which allows looking through the cloud cover to map the effect of the flood~\cite{wagner2024globalflood}.

\begin{figure}[t]
    \centering
    \includegraphics[width=0.9\linewidth]{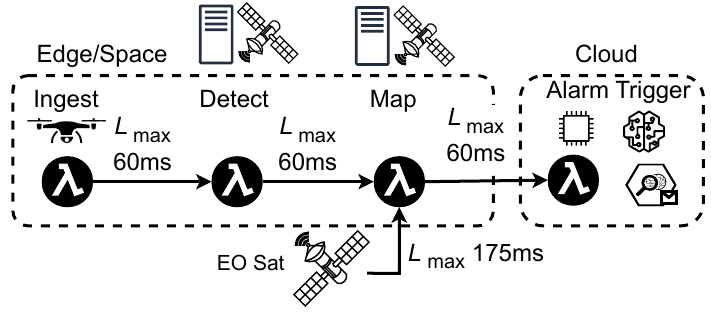}
    \caption{Simplified Serverless Workflow for Flood Disaster Detection}
    \label{fig:use-case-workflow}
\end{figure}

The Flood Disaster detection workflow consists of four serverless functions: Ingest, Detect, Map, and Alarm Trigger, as shown in \cref{fig:use-case-workflow}.
Ingest receives a one-minute sequence of video recorded at 15~frames per second from a drone, along with positional data.
It filters out any unusable frames, e.g., blurry frames or frames recorded while passing through a cloud.
The remaining frames are analyzed by the Detect function using a DNN~\cite{SavingLivesFromAbove2023} to identify people in distress.
The results are forwarded to the Map function, which also receives SAR data from an EO satellite via ISL.
The SAR data is used to map the flooding using a CNN~\cite{FloodCNN_Classification2023} to identify areas that need further investigation by drones.
Finally, the Alarm function accumulates all information and notifies rescue teams about people in need of rescue.
Each function produces a state that serves as input to the next function.
To ensure the timely triggering of alarms and prevent queuing, there is a 60~ms max handoff latency between each function and its successor, which includes all data transfer that needs to take place.

Typically, remote sensing systems, such as JRC’s Global Flood Monitoring, exhibit 8–12h latency, while Copernicus Emergency Management Service faces even longer delays, exceeding 24h when including revisit and processing times, limiting their usefulness for urgent disaster response~\cite{wagner2024globalflood}. LEO satellites offer low-latency connections, even in remote areas, which makes them useful for offloading computing for scientific missions in remote regions~\cite{MobileEdgeCompLEO_AntColony2022} or for people living in poorly connected areas.
LEO satellites also offer significant benefits for processing data from EO satellites.
EO data is very large, e.g., 1.5~TB per day for each Sentinel-2 satellite~\cite{Sentinel2CLaunched2024}.
Since atmospheric interference causes satellite-to-ground bandwidth to be lower around 300~Mbps~\cite{EDRS_Overview} and unreliable~\cite{OEC_ML} compared to 100~Gbps bandwidth of ISL~\cite{OECSurvey2023,DelayNoOption}, processing EO data in space can save time and provide the final results faster. Therefore, Databelt enables functions placed in LEO to run AI models to map floods by analyzing SAR data from EO satellites, which provides all-weather, cloud-penetrating observations, making it ideal for flood mapping~\cite{Singha2024,UrbanSARFloods}.

Due to the fast-moving nature of LEO satellites, selecting suitable state storage nodes and proactively transferring data to the nodes that will execute the functions is essential to fulfilling the SLO.
The data transfer comes with two particular challenges.
The first challenge is to ensure that as little time as possible is wasted with the transfer of the video frames from one function to the next.
The second challenge is prefetching the data from the EO satellite to the LEO satellite node that will execute the Map function and preloading the aggregated flood disaster state used by the Alarm function to the node that will execute this function.
The flood disaster state is stored in the closest Cloud data center because, on a satellite, it would inevitably move at some point to the other side of the Earth, which would increase latency.
The observation data from the EO satellite is continuously being acquired, so it is best obtained directly from that satellite.
Both the observation data from the EO satellite and the alarm state need to be proactively transferred to the satellite nodes that are going to host the Detect and Map functions to enable SLO-compliant execution of the workflow.

\subsection{Scheduling in the 3D~Continuum}

Databelt relies on our previous work, HyperDrive~\cite{HyperDrive2024}, for scheduling serverless functions on the nodes of the 3D~Continuum.
HyperDrive is aware of the end-to-end response time  Service
Level Objectives (SLOs) and is designed to address the challenges introduced by satellites and by the scale of the 3D~Continuum.
Each function of a workflow enters the scheduling pipeline independently when the orchestrator (e.g., Kubernetes) decides that it is ready for scheduling. The scheduler is aware of the task’s dependencies and uses a scheduling context that includes workflow information and a network graph representing the current state of the 3D~Continuum. This enables HyperDrive to consider network quality and ensure response time SLO adherence when making placement decisions. The scheduler is designed as a distinct component that can work in a centralized or distributed mode. It interacts with orchestrators (e.g., Kubernetes) through API calls for sampling nodes, getting network quality of service (QoS) parameters, and committing scheduling decisions. 
None of these operations requires resources to be locked beyond the duration of the API call, which allows multiple scheduler instances to run simultaneously.
To facilitate tracking of information related to a workflow instance, all tasks of a particular workflow instance are handled by the same scheduler instance.

Three important features of HyperDrive that address the challenges of the 3D~Continuum are vicinity selection, network QoS awareness, and satellite temperature awareness.
To preselect a set of candidate nodes for hosting a function, the scheduler samples nodes in a user-defined radius around the predecessor function.
Each of these sampled nodes in the vicinity of the predecessor is further evaluated for resource requirements and whether the network path to it fulfills the latency and bandwidth SLOs.
All nodes that fulfill the resource and network QoS requirements are assessed for temperature limits, if they are satellites.
Specifically, the scheduler checks if a function would cause a satellite to exceed its maximum allowed temperature given its exposure to the sun and the heat generated from the workload.
The nodes are also scored according to their network latencies to ensure that not only can the SLO be fulfilled, but that the fastest possible node can be selected.


\section{Formalized Requirements Model for the 3D Continuum \& Databelt Architecture Overview} \label{sec:architecture}

This section builds upon the functional requirements presented in our previous work ~\cite{HyperDrive2024} and elaborates them into formal model requirements. Additionally, we present the Databelt architecture, describing its core components, including the Databelt Service, Middleware, Storage, Runtime, Controller, and Ingress.

\subsection{Formalized Requirements Model for the 3D Continuum}
The 3D~Continuum presents different challenges not present in the traditional Edge-Cloud Continuum. To enable seamless and efficient execution in the 3D~Continuum, Databelt must account for both environmental and computational characteristics, including dynamic node availability, constrained resources, and orbital movements. Building on the core requirements identified in ~\cite{HyperDrive2024}, we formalize these requirements. 

\subsubsection{Serverless Workflow and Network Model}

The serverless workflow and network topology are the foundation models that integrate the infrastructure with the application properties deployed. 

\paragraph{Serverless Workflow} Serverless workflows consist of interconnected functions that execute independently and typically rely on external state management. Therefore, we model these workflows as a directed acyclic graph (DAG). Let a Serverless Workflow be represented as a DAG $\mathcal{W} = (\mathcal{F}, \mathcal{E})$, where $\mathcal{F} $ is the set of serverless functions, with each node $ f \in \mathcal{F} $ representing a function. $ \mathcal{E} $ is the set of directed edges, where $ (f_i, f_j) \in \mathcal{E} $ represents a communication such that $ f_i $'s output state is required as input for $ f_j $.

\paragraph{Network Topology} To minimize function state access, we must identify the network topology, including its different node types (i.e., edge, cloud, and space) and their characteristics such as bandwidth and latency. Let the Network Topology be represented as a graph $ \mathcal{G} = (\mathcal{N}, \mathcal{L})$, where $ \mathcal{N} $ is the set of nodes, including $\mathcal{N}_C $ (cloud nodes), $\mathcal{N}_E $ (edge nodes), and $ \mathcal{N}_S $ (satellite nodes). $ \mathcal{L}(n_s, n_d)$ is the communication latency between nodes $n_s$ source and $n_d$ destination.

\subsubsection{Formalized Requirements}

To ensure the efficient execution of serverless workflows in the 3D Continuum, we define a set of requirements that capture its unique operational and environmental characteristics.
These requirements provide a foundation for building intelligent mechanisms such as Databelt function state propagation (described in \cref{subsec:state_propagation}).

\paragraph{R-1: Resource Capacity} It ensures that each node has sufficient resources to host the functions assigned to it, maintaining system stability and balanced resource utilization. Let $D_i$ denote the resource demand of function $f_i$, $x_{i,n} \in \{0,1\}$ indicate whether $f_i$ is placed on node $n$, and $R_n$ be the available resource capacity of node $n$.  
The total demand of functions placed on a node must not exceed its available resources:

\begin{equation}
    \sum_{i \in \mathcal{F}} D_{i} \cdot x_{i,n} \leq R_n \quad \forall n \in \mathcal{N}
\end{equation}

\begin{table}[t!]
\centering
\label{tab:databelt-nomenclature}
\resizebox{\linewidth}{!}{%
\begin{tabular}{|p{1.75cm}@{\hskip 4pt}p{9.5cm}|}
\hline
& \\[-1ex]
\multicolumn{2}{|l|}{\textbf{Nomenclature}} \\[0.5ex]
& \\[-1.5ex]
\multicolumn{2}{|l|}{\textit{Mathematical Notation and Parameters}} \\[0.5ex]
& \\[-1.5ex]
$\mathcal{W} = (\mathcal{F}, \mathcal{E})$ & Serverless workflow DAG, where $\mathcal{F}$ is the set of functions and $\mathcal{E}$ is the set of directed edges. \\[0.5ex]
$\mathcal{G} = (\mathcal{N}, \mathcal{L})$ & Network topology graph, where $\mathcal{N}$ is the set of nodes and $\mathcal{L}$ is the latency between them. \\[0.5ex]
$(f_i, f_j) \in \mathcal{E}$ & Directed edge representing dependency: $f_i$'s output is required by $f_j$. \\[0.5ex]
$\mathcal{L}(n_s, n_d)$ & Latency between source node $n_s$ and destination node $n_d$. \\[0.5ex]
$x_{i,n}$ & Binary placement variable: 1 if function $f_i$ is placed on node $n$, 0 otherwise. \\[0.5ex]
$D_i$ & Resource demand (e.g., CPU, memory) of function $f_i$. \\[0.5ex]
$R_n$ & Available resource capacity at node $n$. \\[0.5ex]
$T_{\text{orb}}^n$ & Baseline orbital temperature of node $n$. \\[0.5ex]
$T_{\text{exc}}^{in}$ & Temperature increase on node $n$ due to execution of function $f_i$. \\[0.5ex]
$T_{\text{max}}^n$ & Maximum allowed operational temperature of node $n$. \\[0.5ex]
$S_{ij}$ & SLO latency threshold between function pair $(f_i, f_j)$. \\[0.5ex]
$\gamma(n_s, n_d)$ & Penalty coefficient for state propagation from node $n_s$ to $n_d$, 0 if $n_s = n_d$. \\[0.5ex]
$P_i$ & Power demand of function $f_i$. \\[0.5ex]
$P_{\text{avail}}^n$ & Available power capacity at node $n$. \\[0.5ex]
$a_n(t)$ & Availability indicator of node $n$ at time $t$, 1 if available, 0 otherwise. \\[0.5ex]
$A(t)$ & Set of available nodes at time $t$ that satisfy connectivity, and SLOs. \\[0.5ex]
$\mathcal{T}$ & Set of required node types for satellite availability (e.g., drone, EO satellite, ground station). \\[0.5ex]
$r_{\tau}(n,t)$ & Binary reachability indicator: 1 if node $n$ can reach a node of type $\tau$ at time $t$, 0 otherwise. \\[0.5ex]
\hline
\end{tabular}
}
\end{table}

\paragraph{R-2: Temperature Limit} Increasing computing demand in orbit exacerbates thermal management challenges, as satellites must rely on radiation to dissipate heat in space. This issue is critical for LEO satellites, which operate with fixed hardware that cannot be upgraded or repaired after launch. Exposed to extreme temperatures (from $-120^{\circ}$C in shadow to $+120^{\circ}$C in sunlight~\cite{SpaceEnvEffectsNASA2020}), these satellites must maintain performance despite harsh thermal cycles. Without mitigation, prolonged exposure to heat degrades key components such as the CPU~\cite{Vision6G2024,SCCaseStudy2023,SatelliteComputingVision2023,SkyEdge2021}. Thus, the temperature of each node $n$ in \( \mathcal{N} \) must not exceed its maximum allowed temperature $T_{\text{max}}$, considering the maximum temperature caused by the satellite exposure to the sun and the temperature sum increase due to the execution of the each function $T_{\text{exc}}$:

\begin{equation}
    T_{\text{orb}}^{n} + \sum_{i \in \mathcal{F}} T_{\text{exc}}^{in} \leq T_{\text{max}}^{n} \quad \forall n \in \mathcal{N}
\end{equation}

\paragraph{R-3: Energy Capacity} As computational demands grow on satellites, managing energy consumption becomes critical due to strict size and weight limitations~\cite{Vision6G2024}. 
Since satellites may not fully recharge their batteries during each orbital daylight period, this constraint ensures a balance between energy consumption and energy harvesting to avoid battery depletion.
Thus, $P_i$ denotes the power demand of function $f_i$, $x_{i,n} \in \{0,1\}$ indicates whether function $f_i$ is placed on node $n$, and $P_{\text{avail}}^n$ represents the available power at node $n$. The total power consumption of functions placed on a node must not exceed its available power:

\begin{equation}
    \sum_{i \in \mathcal{F}} P_{i} \cdot x_{i,n} \leq P_{\text{avail}}^n \quad \forall n \in \mathcal{N}
\end{equation}

\paragraph{R-4: SLOs} Typically, to ensure performance, serverless workflows must satisfy specific SLOs. Meeting SLOs requirements in orbit is challenging due to the network specifics, orbital movements, battery, and heat conditions of satellites~\cite{ResourceAllocationLEOSurvey}. This constraint ensures that the data does not exceed certain predefined requirements. In our scenario, the function execution under SLO requirements ensures a fast reaction in a disaster response scenario.  This means that communication between functions must meet performance criteria defined by the user to minimize delays.  Thus, the latency SLOs $S_{ij}$ must be met for each function invocation pair for each function pair $ (f_i, f_j) \in \mathcal{E}$. The latency $ \mathcal{L}(n_s, n_d)$ of the path between nodes  $ n_s $ and $ n_d $ in \( \mathcal{N} \) must not exceed the SLO $S{_ij}$:

\begin{equation}
    \mathcal{L}(n_s, n_d) \leq S_{ij} \quad \forall (f_i, f_j) \in \mathcal{E}, \forall (n_s, n_d) \in \mathcal{N}
\end{equation}

\paragraph{R-5: Satellite Node Availability} 
Due to their continuous orbital movement, satellite nodes $\mathcal{N}$ exhibit dynamic availability over time~\cite{GearingUp2018}. A satellite node $n$ is considered available at time $t$ if it can simultaneously connect to a drone, an EO satellite, and a ground station, directly or through ISL, with latency and bandwidth that meet communication thresholds~\cite{DelayNoOption, AnalysisISL2021}.  
Thus, the availability of a satellite node $n \in \mathcal{N}$ at time $t$ depends on its connectivity to required node types $\mathcal{T}$ (e.g., drone, EO satellite, ground station). We define the set of available nodes at time $t$, $A(t)$, where a node $n \in \mathcal{N}$ is included if it can reach all required node types $\tau \in \mathcal{T}$, either directly or via ISLs. The binary function $r_{\tau}(n,t) \in \{0,1\}$ indicates if node $n$ can reach a node of type $\tau$ at time $t$. Hence, the satellite node availability can be defined as follows:

\begin{equation}
    A(t) = \{ n \in \mathcal{N} \mid \bigwedge_{\tau \in \mathcal{T}} r_{\tau}(n,t) = 1 \}, \quad r_{\tau}(n,t) \in \{0,1\}
\end{equation}

\paragraph{R-6: Function Placement} Functions within a serverless workflow often depend on an intermediate state for execution. However, retrieving the state from remote nodes may lead to network latency and congestion, degrading performance. In the dynamic 3D Continuum, multi-hop state access can significantly increase workflow latency. Therefore, we define that each function instance $f_i$ in the workflow must be placed on a node $n$, and its successor $f_j$ is placed on one node $n'$, then Databelt can find the shortest path between these two nodes. Moreover, function placement is restricted to nodes that are available at execution time. Thus, we define $x_{i,n} \in \{0,1\}$ as a binary variable indicating whether function $f_i$ is placed on node $n$, where placement is restricted to the available nodes $n \in A(t)$ at execution time. It is defined as follows:

\begin{equation}
    \sum_{n \in A(t)} x_{i,n} = 1 \quad \forall f_i \in \mathcal{F}
\end{equation}

\paragraph{R-7: Data Locality} 
To minimize latency and reduce dependency on external storage services, functions should access the state from the same execution node whenever possible. However, given that the target function is not scheduled at the time when the state propagation happens, execution on the same node cannot always be guaranteed. Therefore, we introduce a locality-aware penalty mechanism that discourages remote state propagation while allowing strategic intermediate state placements when necessary. Local state propagation (i.e., where $ n_s = n_d $) remains preferred, but state access from nearby nodes incurs a penalty based on network distance and latency, where $ \gamma(n_s, n_d) $ is a penalty coefficient that increases with the network distance between nodes $ n_s $ and $ n_d $. If $ n_s = n_d $, then $ \gamma(n_s, n_d) = 0 $, ensuring no penalty for local execution. The data locality constraint can be expressed as:

\begin{equation}
    \sum_{(f_i, f_j) \in \mathcal{E}} \sum_{(n_s, n_d) \in \mathcal{N}} \gamma(n_s, n_d) \cdot x_{i,s} x_{j,d} \leq \sum_{(f_i, f_j) \in \mathcal{E}} x_{i,s} x_{j,s}
\end{equation}

\subsubsection{Overarching Goal}

The formalized requirements outlined in the previous section reflect the key challenges of executing serverless workflows in the 3D Continuum. To implement these requirements effectively, we formalize them as constraints that define the conditions under which a function can be executed. These constraints ensure that the workflows remain feasible under environmental and infrastructure limitations of the 3D Continuum. Therefore, Databelt's goal is to minimize overall execution time while adhering to these constraints.

\paragraph{Objective} To decrease the overall execution time, we must minimize the state propagation latency between functions. Considering a workflow $ \mathcal{W} $, where $ \mathcal{L}(n_s, n_d) $ is the latency between nodes $ n_s $ and $ n_d $,  $ x_{i,s} $ is a binary variable indicating whether function $ f_i $ is placed on node $ n_s $, and  $ x_{j,d} $ is a binary variable indicating whether function $ f_j $ is placed on node $ n_d $. Thus, the objective function can be defined as:

\begin{equation}
    \min_{x} \sum_{(f_i, f_j) \in \mathcal{E}} \sum_{(n_s, n_d) \in \mathcal{N}} \mathcal{L}(n_s, n_d) \cdot x_{i,s} x_{j,d}
\end{equation}

\paragraph{Optimization Problem}  In a mixed environment, Serverless workflows can be executed on ground-based or LEO Edge nodes. Thus, we need to take into account the workflow composition to identify the dependencies and interactions between the functions. 
Our heuristic approach aims to minimize state propagation latency while adhering to SLO constraints. Hence, the final optimization problem can be expressed as:
\begin{equation}
\begin{aligned}
    & \min_{x} \sum_{(f_i, f_j) \in \mathcal{E}} \sum_{(n_s, n_d) \in \mathcal{N}}  \left( \mathcal{L}(n_s, n_d) + \gamma(n_s, n_d) \right) \cdot x_{i,s} x_{j,d} \\
    \text{s.t.} \quad 
    \quad & \sum_{i \in \mathcal{F}} D_{i} \cdot x_{i,n} \leq R_n \quad \forall n \in \mathcal{N} \\
    & T_{\text{orb}}^{n}(t_i) + \sum_{i \in \mathcal{F}} T_{\text{exc}}^{in} \leq T_{\text{max}}^{n} \quad \forall n \in \mathcal{N} \\
    &  \sum_{i \in \mathcal{F}} P_{i} \cdot x_{i,n} \leq P_{\text{avail}}^n \quad \forall n \in \mathcal{N} \\
    & \mathcal{L}(n_s, n_d) \leq S_{ij} \quad \forall (f_i, f_j) \in \mathcal{E}, \forall (n_s, n_d) \in \mathcal{N} \\ 
    & \sum_{n \in A(t)} x_{i,n} = 1 \quad \forall f_i \in \mathcal{F} \\
    & \sum_{(f_i, f_j) \in \mathcal{E}} \sum_{(n_s, n_d) \in \mathcal{N}} \gamma(n_s, n_d) \cdot x_{i,s} x_{j,d} \leq \sum_{(f_i, f_j) \in \mathcal{E}} x_{i,s} x_{j,s} \\
    & x \in \{0,1\} \quad \forall f_i \in \mathcal{F}, \forall n \in \mathcal{N}
\end{aligned}
\end{equation}

The Databelt serverless workflow model aims to optimize serverless execution across the 3D Continuum. By incorporating an SLO-aware network topology, Databelt dynamically positions function states to reduce latency and improve the overall workflow execution. The Databelt model incorporates the environmental properties from the different layers, such as satellite orbit movements, and power consumption to proactively migrate function states based on network conditions and SLO constraints.


\subsection{Databelt Architecture Overview}

\begin{figure}[t]
    \centering
    \includegraphics[width=\linewidth]{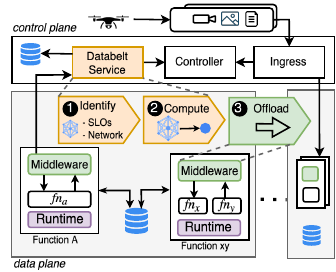}
    \caption{Databelt Architecture Overview showing its three-phase execution: Identify, Compute, and Offload.}
    \label{fig:architecture_overview}
\end{figure}

Databelt operates in three phases (shown in \cref{fig:architecture_overview}) designed to optimize the propagation of function states across the 3D~Continuum: Identify, Compute, and Offload. These phases, detailed in \cref{subsec:state_propagation}, determine how function states are dynamically allocated based on network conditions, node availability, and predefined SLOs. 
Databelt's architecture consists of two planes: the data plane, which coordinates and executes each function, and the control plane, which manages system-wide coordination. 

The control plane is responsible for system-wide decisions such as registering new nodes, scheduling decisions, SLO enforcement, and routing initial requests. It includes the Databelt Service, (Global) Storage, Controller, and Ingress. In contrast, the data plane comprises Middleware, Runtime, and (Local) Storage, which are crucial for executing functions, as they manage the function lifecycle, state access, and data propagation.

\subsubsection{Databelt Components}
Databelt is composed of six components: Databelt Service, Middleware, Storage, Runtime, Controller, Storage and Ingress.  

\paragraph{Databelt Service} The Databelt Service is designed to be asynchronous, allowing parallel handling of multiple workflows without introducing control-plane contention. It operates in the control plane and is responsible for topology management and state computation across the 3D continuum. The service maintains an internal topology representation by periodically querying the orchestrator API and collecting information on node latency, bandwidth, and SLO constraints. Based on this real-time network view, it dynamically determines optimal state placement by selecting target nodes that minimize propagation latency while complying with SLO requirements. Moreover, the Databelt Service provides an API interface to enable the Middleware to retrieve decisions related to the state migration. 

\paragraph{Middleware} It enables transparent state access and state propagation. It retrieves execution metadata and communicates with the Databelt Service to make real-time decisions on state migration. The middleware retrieves the propagation information from the Databelt services and offloads the state in the specified storage node. The Databelt middleware also enables low-latency function state sharing within functions in the same sandbox; it gathers information from the function runtime to identify whether there is one or multiple functions. Then, it merges the functions' state within one single request to reduce redundant storage operations. 

\paragraph{Storage} Databelt uses a two-layer storage with local and global storage to minimize latency during function state access. Local storage ensures that function states are readily available at the execution node. In contrast, global storage provides redundancy, allowing functions to retrieve the required state even if the local copy becomes unavailable.

\paragraph{Runtime} It manages the function lifecycle. The serverless runtime is triggered by the controller during provisioning and executes the corresponding function code. Databelt leverages existing techniques~\cite{Cwasi2023,2024fusionize,diffuse,Fusion} to enable a shared serverless runtime across trusted functions within a workflow, also known as function fusion. The runtime identifies eligible fusion groups by detecting functions that are co-located. It checks whether the required functions are nearby and suitable for fusion based on function and workflow properties, developer annotations, or container image analysis. Additionally, when a request arrives, the Databelt runtime checks whether the invoked function is part of a fused group. If so, instead of delegating each function to retrieve its state, Databelt preloads all fused states simultaneously, reducing latency, communication overhead, and storage operations. \cref{subsec:function_fusion} describes the function state fusion in detail.

\paragraph{Controller} It is responsible for system-wide decision-making and provisioning tasks. The controller manages function scheduling across nodes in the cluster and monitors each node's resource utilization while enforcing user-defined SLO requirements. Additionally, the controller dynamically adjusts the number of function instances based on workload demands, scaling functions up or down as needed, ensuring efficient execution, resource optimization, and resilient workflow operation across the 3D Continuum.

\paragraph{Ingress} It serves as the system’s entry point, receiving incoming function events and routing them to the appropriate execution nodes. It is responsible for handling initial requests, load balancing across nodes, and efficiently forwarding requests to functions that are already running or triggering the controller if no function is running.


\section{Databelt Main Runtime Mechanisms} \label{sec:mechanisms}
In this section, we introduce Databelt's state migration and function state fusion mechanisms, which optimize execution by dynamically allocating state and co-locating related function states to minimize latency and redundant storage operations.


\subsection{Function State Propagation} \label{subsec:state_propagation}
Databelt's function state propagation (shown in \cref{fig:architecture_overview}) happens in three phases: \emph{Identify}, which collects metadata from the nodes across the 3D continuum, \emph{Compute} selects the most appropriate set of nodes, and calculates the shortest path between them, and \emph{Offload}, which proactively places the function state to the best candidate. To minimize runtime overhead, Databelt decouples node identification (\emph{Identify}) and placement computation (\emph{Compute}) from function execution (\emph{Offload}). All propagation-related decisions, including candidate node selection, SLO filtering, and routing, are precomputed in the control plane by the Databelt Service, which maintains a global view of the system topology. At runtime, functions retrieve these precomputed decisions via lightweight API calls, incurring negligible computational or communication overhead. This control–data plane separation ensures that Databelt’s mechanisms do not interfere with function execution time.

\paragraph{Identify} It constructs a real-time topology graph of all available execution nodes across the 3D Continuum. Nodes register with the Databelt Service, providing metadata such as bandwidth, latency, position, and current state availability, enabling continuous network profiling. Moreover, Databelt identifies SLO latency constraints defined by the user during deployment and maps these constraints to network conditions, such as link latency and available bandwidth, to define feasible execution paths for functions. Based on function dependencies, Databelt determines a set of candidate nodes for state placement, ensuring optimal function execution while minimizing propagation overhead. As detailed in \cref{alg:identify}, Databelt filters the set of nodes $\mathcal{N}$ to retain only those that are available at the current time $t$ (line~\ref{line:node_filter}). Edges are pruned to preserve only links between those nodes available (line~\ref{line:edges}). If so, the corresponding edge is added to the pruned graph along with its associated latency $\mathcal{L}(n_s,n_d)$ and bandwidth $b$ (line~\ref{line:graph}). Finally, the identify phase returns the pruned topology $N_t = (V_N, E_N)$ (line~\ref{line:output}), which captures only currently feasible nodes and edges for function state propagation.

\begin{algorithm}[t]
\caption{Databelt Function State Propagation: Identify}
\label{alg:identify}
\begin{algorithmic}[1]
    \Statex \textbf{Input: } $\mathcal{N}$: set of nodes; 
    \Statex $\mathcal{L}$: set of links $(n_s,n_d,l,b)$;
    \Statex $t$: current time
    \Statex \textbf{Output: } $N = (V_N, E_N)$: pruned network graph at time $t$.

   \State $V_N \gets \{\, n \in \mathcal{N} \mid a_n(t) = 1 \,\}$ \Comment{keep only available nodes} \label{line:node_filter}
    \State $E_N \gets \emptyset$
    \ForAll{$(n_s, n_d, l, b) \in \mathcal{L}$} 
        \If{$n_s \in V_N$ \textbf{and} $n_d \in V_N$} \label{line:edges}
            \State $E_N \gets E_N \cup \{ (n_s, n_d;\ l=\mathcal{L}(n_s,n_d),\ bw=b) \}$ \label{line:graph}
        \EndIf
    \EndFor
    \State \Return $N = (V_N, E_N)$ \label{line:output}
\end{algorithmic}
\end{algorithm}

\paragraph{Compute} 
To ensure scalability in large topologies, Databelt applies a topology-aware pruning, which limits candidate node selection based on communication paths and SLO constraints, significantly reducing the decision space, and provides a service interface, enabling horizontal scaling of the Databelt Service. This ensures that the control plane does not become a bottleneck as the number of workflows, nodes, or function invocations increases.
Databelt Service determines the shortest propagation path by selecting the lowest-latency route for intermediate state transfer between serverless functions within the workflow. In a workflow consisting of multiple functions distributed across edge, space, and cloud environments, Databelt identifies an intermediate subset of nodes for the state placement. Since the final function is likely to be executed in the cloud due to high resource demands not available in satellites (e.g., AI inference), function states are strategically distributed among these intermediate subsets of nodes. Even if a function is not executed on a specific node where the state is present at runtime, it can access the necessary state from a nearby location, minimizing propagation and state access latency.

\begin{figure}[t]
    \centering
    \includegraphics[width=\linewidth]{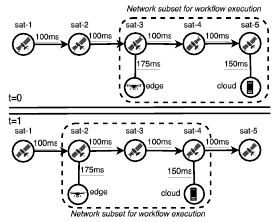}
    \caption{Databelt state propagation adapting the network subset for state placement to the satellite orbit movement and workflow execution at instant $t=0$ and $t=1$.}
    \label{fig:propagation}
\end{figure}

\begin{algorithm}[!t]
\caption{Databelt Function State Propagation: Compute}
\label{alg:propagate-policy}
\begin{algorithmic}[1]
    \Statex \textbf{Input: } $N = (V_N, E_N)$: Network graph;
    \Statex $|k|$: Data size;
    \Statex $b$: Bandwidth;
    \Statex $n_D$: Destination node;
    \Statex $t_{max}$: Maximum allowed migration time;
    \Statex \textbf{Output: } Selected node $n_C$ for data propagation;

    \State $n_S \gets$ \Call{GetHostNode}{$N$} \Comment{Source node} \label{line:getHost}
    \State $P \gets$ \Call{Dijkstra}{N, $n_S$, $n_D$} \Comment{Shortest path from $n_S$ to $n_D$} \label{line:Dijkstra}
    \State $P \gets$ \Call{Reverse}{P} \Comment{Reverse path for evaluation} \label{line:reverse}
    \ForAll{$(n_C, l_C) \in P$}
        \State $t_{mig} \gets l_C + \frac{|k|}{b} + l_C$ \Comment{Migration time calculation} \label{line:migration}
        \If{$t_{mig} > t_{max}$}
            \State \textbf{continue} \Comment{Skip if migration time exceeds maximum}
        \EndIf
        \State \Return $n_C$ \Comment{Return suitable candidate node} \label{line:node_found}
    \EndFor
    \State \Return $n_S$ \Comment{Fallback to source node if no candidates found} \label{line:node_notfound}
\end{algorithmic}
\end{algorithm}

At runtime, when a workflow is triggered (e.g., by an edge device at instant $t=0$), the initial function $f_1$ processes the data and requests Databelt to determine the optimal state placement. 
Databelt selects nodes that satisfy SLO constraints and anticipates network topology changes (\cref{fig:propagation}), such as the orbital movement of satellites. To avoid potential failures, it excludes nodes likely to move out of communication range (e.g., sat-5), ensuring continuous accessibility of function states throughout the workflow execution.
To optimize inter-function communication, Databelt computes the shortest propagation path between execution nodes. Specifically, it uses Dijkstra’s algorithm (line~\ref{line:Dijkstra}) to identify the lowest-latency route from the current execution node ($n_S$) to the final destination node ($n_D$) in the cloud. As detailed in \cref{alg:propagate-policy}, the computed path is then reversed (line~\ref{line:reverse}) to prioritize nodes closer to the destination, facilitating more efficient intermediate state placement.
For each candidate node $n_C$ along the reversed path, Databelt estimates the total migration time $t_{mig}$ based on link latency $l_C$, data size $|k|$, and available bandwidth $b$ (line~\ref{line:migration}). The first node satisfying the maximum migration time constraint $t_{max}$ is selected for state placement (line~\ref{line:node_found}). If no candidate fulfills the SLO requirements, the source node $n_S$ is retained for storing the state (line~\ref{line:node_notfound}).

\begin{algorithm}[t]
\caption{Databelt Function State Propagation: Offload}
\label{alg:offload}
\begin{algorithmic}[1]
    \Statex \textbf{Input: } $N=(V_N,E_N)$: network graph; 
    \Statex $f_i$: current function; 
    \Statex $s_i$: state produced by $f_i$; 
    \Statex $t$: current time
    \Statex \textbf{Output: } Final placement node $n_{\text{place}}$ for $s_i$

    \State $n_S \gets \Call{GetHostNode}{N}$ \Comment{executor of $f_i$}
    \State $n_C \gets \Call{GetPlacementDecision}{f_i}$ \Comment{target from control plane (Compute)} \label{line:get_node}
    \If{$a_{n_C}(t) = 1$}  \label{line:node_avail}
        \State \Call{Place}{$s_i, n_C$} \label{line:transfer}
        \State $n_{\text{place}} \gets n_C$
    \Else
        \State \Call{Place}{$s_i, n_S$} \Comment{fallback if target currently unavailable}\label{line:local}
        \State $n_{\text{place}} \gets n_S$
    \EndIf
    \State \Return $n_{\text{place}}$\label{line:target}
\end{algorithmic}
\end{algorithm}

\paragraph{Offload} The state transfer follows the shortest propagation path computed during the Compute phase, ensuring minimal communication overhead. Upon completion of the function, Databelt proactively releases the generated state to a pre-selected target node, strategically positioning it near the expected execution location of the next function. This proactive placement reduces remote data fetches, optimizes inter-function communication, and maintains compliance with predefined SLO constraints.
At instant $t=1$, the network topology changes as satellites continue their orbital movement (\cref{fig:propagation}). However, because Databelt anticipates these changes during the Identify and Compute phases, the subsequent function $f_2$ can seamlessly retrieve the prepositioned state from nodes remaining within communication range (e.g., sat-2 to sat-4), preserving low retrieval latency and workflow continuity.
By continuously adapting state placement to dynamic network conditions, Databelt minimizes end-to-end execution time while ensuring ongoing SLO compliance. As detailed in \cref{alg:offload}, when a function $f_i$ completes, the produced state $s_i$ is proactively placed on the pre-selected target node $n_C$ determined in the Compute phase (line ~\ref{line:get_node}). If the target node is available at time $t$ (line ~\ref{line:node_avail}), the state is transferred there (line ~\ref{line:transfer}). Otherwise, Databelt falls back to storing the state locally at $n_S$ (line ~\ref{line:local}). The algorithm then returns the final placement node (line ~\ref{line:target}), which is then forwarded to the target function.

\subsection{Function State Fusion}\label{subsec:function_fusion}

Function fusion refers to the co-location of multiple functions within the same sandbox execution environment, as shown in \cref{fig:architecture_overview}, where $fn_x$ and $fn_y$ share the same function runtime. Databelt introduces a middleware that operates as a communication layer within each function sandbox. When a function scales, the runtime identifies which functions can be fused. Upon receiving a request, the middleware retrieves the required states for the fused functions from local or global storage, reducing the number of I/O operations.

\begin{figure}[t]
   \centering
   \resizebox{\linewidth}{!}{%
    \begin{tikzpicture}
        \node[anchor=west, text=teal] at (0, 0) {9eed920b-1680-461e-ae21:};
        \node[anchor=west, text=violet] at (3.9, 0) {2001:db8::1};
        \node[anchor=west, text=blue] at (5.6, 0) {:d4aa0228-ff89-43ad-8934};
        
        \draw[thick, teal] (0.2,-0.3) -- (0.2,-0.6) -- (3.8,-0.6) -- (3.8,-0.3);
        \draw[thick, violet] (4.1,-0.3) -- (4.1,-0.6) -- (5.7,-0.6) -- (5.7,-0.3);
        \draw[thick, blue] (5.9,-0.3) -- (5.9,-0.6) -- (9.4,-0.6) -- (9.4,-0.3);
        
        \node[anchor=north, text=teal] at (2,-0.6) {\small Workflow ID};
        \node[anchor=north, text=violet] at (4.9,-0.6) {\small Storage Address};
        \node[anchor=north, text=blue] at (7.6,-0.6) {\small Function ID};
    \end{tikzpicture}
    }
    \caption{Databelt Function State Key, used by functions to locate and retrieve state from nearby storage nodes.}
    \label{fig:state-key}
\end{figure}
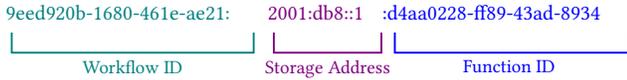

Each function state is identified by a unique workflow-specific key, allowing functions in a workflow to locate and retrieve the required state. When a function executes, it receives a Databelt State Key as input, and at the end of execution, the function generates a new state key and propagates it as the function output. Each state object consists of three parts, as shown in \cref{fig:state-key}, forming a unique state identifier: WorkflowID, which identifies the workflow instance; Storage Address, which specifies where the state is stored; and FunctionID, which uniquely references the function instance. 
Although fused function states are retrieved together to minimize redundant storage operations, Databelt ensures that each function only accesses the subset of state explicitly passed via its Databelt State Key. This key-based isolation guarantees that the function state is not shared indiscriminately across functions, even when co-located. Updates are propagated only when a function completes and writes its output state. Moreover, because Databelt treats state objects as immutable within each invocation, differences in update frequency across functions do not lead to inconsistencies. Databelt is not designed for workflows that require synchronization across functional instances. Such applications typically assume consistent state enforcement, which is outside the scope of Databelt.

\begin{figure}[t]
    \centering
    \includegraphics[width=0.8\linewidth]{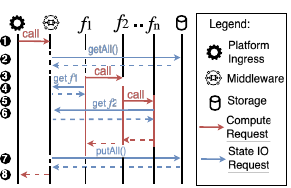}
    \caption{Databelt function state fusion, where Databelt middleware is responsible for the state operation management.}
    \label{fig:function_fusion}
\end{figure}

\cref{fig:function_fusion} shows how Databelt components interact with the functions and storage to retrieve the function state from fused functions. More specifically, \numbercircle{1} an event request triggers the function scale-up process from the serverless platform. Once the serverless platform starts the function runtime, the runtime decides which functions to fuse. Once the function is initialized, the request is forwarded to the function middleware. \numbercircle{2} The middleware gathers information from the runtime related to the fused function to identify which states should be retrieved. If the required states are in the local storage, the middleware fetches them directly from there to minimize latency. Otherwise, it retrieves the states from global storage, ensuring all necessary data is available for the execution of the fused functions. In \numbercircle{3} $f_1$ starts computing the request. In \numbercircle{4}, $f_1$ retrieves its state directly from the middleware, as the data is already present in the process, avoiding additional external requests. In \numbercircle{5} $f_1$ calls $f_2$ to continue the function execution. In \numbercircle{6}, $f_2$ state is already in the process. Thus, $f_2$ only needs to retrieve its state directly from the middleware. In \numbercircle{7}, once every fused function finishes, the middleware merges the state and propagates it to the storage. Finally, in \numbercircle{8}, the function execution finishes and returns the end result to the client.


\section{Prototype Implementation} \label{sec:impl}

The prototype of the Databelt is available as open-source\footnote{\url{https://github.com/polaris-slo-cloud/databelt}}. 
To ensure lightweight and safety in resource-constrained environments such as the 3D~Continuum, the Databelt state-propagation module is written in Rust and leverages WebAssembly, more specifically, the WasmEdge~\cite{wasmedge} runtime. 

Databelt implements a topology handler that receives as input i) the cluster node graph and ii) the user-defined SLO targets.
To create the node graph, the Databelt service has a thread that continuously communicates with the orchestrator API to gather real-time node information. Moreover, the Databelt service exposes an HTTP API to deliver real-time information to the Databelt middleware. Upon receiving the HTTP requests, the Databelt service finds the next best node on a different thread. The Databelt service relies on Rust libraries \path{tokio} for the HTTP service and Rust \path{Mutex} mechanisms to provide thread safety.

Moreover, to enable state offloading, databelt middleware is imported directly into user functions. It extracts necessary metadata (e.g., function state keys) and communicates with the Databelt Service using Rust \path{reqwest} HTTP client. The middleware provides state management capabilities by leveraging KVS (Redis) deployed on each cluster node. For serialization and deserialization of state data, Databelt middleware uses \path{serde} and \path{serde_json}. For the remaining components (i.e., Controller and Ingress), Databelt relies on state-of-the-art orchestrators such as Kubernetes.


\section{Evaluation} \label{sec:evaluation}


We design experiments to replicate our illustrative scenario (\cref{subsec:scenario}) and we evaluate Databelt through experiments to analyze its state propagation and function fusion mechanisms, detailed in \cref{sec:mechanisms}. These experiments are divided into state propagation performance, state propagation scalability, and function fusion latency. These experiments reflect common serverless invocation patterns, such as sequential and parallel executions~\cite{CloudProgrammingSimplified}. The state propagation experiments measure workflow execution latency and scalability under different configurations and network conditions. To isolate the benefits of the state propagation mechanisms, we use the function state fusion depth of 1. In the function fusion experiments, we focus on the execution latency result after embedding function states. The separation of state propagation and function state fusion experiments allows us to clearly distinguish the impact of each mechanism on the overall workflow performance.

\paragraph{Experimental Workflow} Our experimental workflow includes four serverless functions: Ingest, Detect, Map, and Alarm, as described in \cref{subsec:scenario}. Databelt determines the optimal state and function placement to minimize latency, respecting the user-defined SLO.
The function fusion experiments use configurations where functions retrieve and store states individually or bundled. Each experiment runs with a state stored on local (stateful) or remote storage (stateless).

\paragraph{Objectives} Our experiments measure workflow latency and throughput to evaluate Databelt efficiency under varying loads. We collect state read and write times to evaluate the performance of retrieval and storage operations. We also measure state distance as the mean network hops between the function's execution node and the state storage node to analyze Databelt state propagation efficiency. Additionally, we record local state availability to determine if a function can access the state locally. 

\paragraph{Baselines} We compare Databelt with two baseline storage configurations: Stateless and Random. In both baselines, functions retrieve and store their state individually, without any fusion mechanisms. The baselines are: (a) Stateless, where all state is stored in a global KVS on the cloud node. Each function invocation fetches its own state from the cloud and writes back the updated result independently. (b) Random, where the state is stored randomly across the nodes in the cluster. 
These baselines are intentionally chosen to reflect common state-of-the-art serverless configurations and to isolate the benefits of Databelt's topology-aware propagation.  


\begin{table}[t]
\caption{Testbed specifications with node type, hardware details, quantity (\textbf{Qty}), and simulated network latencies in ms (To Cloud and To Sat).}
\label{tab:testbed-config}
\resizebox{\linewidth}{!}{%
\begin{tabular}{@{}lllllrr@{}}
\toprule
\textbf{Type}       & \textbf{Model}       & \textbf{RAM} & \textbf{CPU (GHz)} & \textbf{Qty} & \textbf{To Cloud} & \textbf{To Sat} \\\midrule
Cloud                    & Raspberry Pi 5B      & 8GB          & 4x2.4              & 1               & -                        & 45 - 75 ms                     \\ 
Satellite                & Raspberry Pi 5B      & 8GB          & 4x2.4              & 3               & 45 - 75 ms                 & 1 - 20 ms               \\ 
Satellite                & Raspberry Pi 4B      & 8GB          & 4x1.8              & 3               & 45 - 75 ms                & 1 - 20 ms               \\ 
Edge                     & Raspberry Pi 4B      & 2GB          & 4x1.5              & 1               & 1 - 20 ms                & 45 - 75 ms                \\ \bottomrule
\end{tabular}%
}
\end{table}

\subsection{Experimental Setup}
Deploying applications in space remains costly and complex. Thus, we built a testbed that mimics the compute and network heterogeneity of the 3D Continuum, inspired by the hardware setups that are available through commercial services such as Loft Orbital’s Virtual Missions~\cite{loftorbital}, which offer space-capable infrastructure.
In our testbed, we execute the designed workflow and evaluate the function state propagation. We deployed a MicroK8s~\cite{microk8s} cluster with eight nodes representing different node types in the 3D Continuum as described in \cref{tab:testbed-config}. Each node is running Ubuntu Server for ARM 24.04 LTS with Redis~\cite{redis} as the KVS. The experimental workflow is deployed as Knative~\cite{knative} services.  We configure periodic link disruptions using cron jobs and \path{tc}~\cite{tc_manpage} to simulate dynamic satellite movement and varying communication paths, simulating satellite-to-satellite and satellite-to-cloud communication, thus mimicking the satellite orbit to get out of range of ground nodes. To avoid bias in the results, all experiments were executed 10 times, and the results shown are the corresponding mean values.

\begin{table}[t]
\caption{Function state propagation results with state fusion depth 1.}
\label{tab:function_state_results}
\resizebox{\columnwidth}{!}{%
\begin{tabular}{ccccccccc}
\toprule
\textbf{Input Size} & \textbf{System} & \textbf{Latency (s)} & \textbf{Read (s)} & \textbf{Write (s)} & \textbf{RPS} & \textbf{SLO Viol. (\%)} & \textbf{CPU (\%)} & \textbf{RAM (MB)} \\
\midrule
\rowcolor[gray]{0.9} 10 & Databelt   & 7.90  & 0.64 & 1.74 & 0.1266 & 0   & 16.4 & 1320 \\
   & Random     & 10.76 & 1.90 & 1.85 & 0.0929 & 100 & 16.4 & 1423 \\
   & Stateless  & 12.47 & 2.43 & 2.07 & 0.0802 & 100 & 16.5 & 1423 \\
\midrule
20 & Databelt   & 13.81 & 1.26 & 3.10 & 0.0724 & 0   & 17.0 & 1324 \\
   & Random     & 18.86 & 3.49 & 3.31 & 0.0530 & 40  & 16.4 & 1423 \\
   & Stateless  & 20.27 & 4.00 & 3.43 & 0.0493 & 80  & 16.3 & 1423 \\
\midrule
\rowcolor[gray]{0.9} 30 & Databelt   & 19.39 & 2.04 & 4.26 & 0.0516 & 0   & 17.3 & 1333 \\
   & Random     & 25.71 & 5.54 & 3.85 & 0.0389 & 30  & 16.3 & 1423 \\
   & Stateless  & 27.94 & 5.67 & 4.69 & 0.0358 & 100 & 16.2 & 1424 \\
\midrule
40 & Databelt   & 24.48 & 2.49 & 5.46 & 0.0409 & 0   & 17.2 & 1339 \\
   & Random     & 29.53 & 5.73 & 5.13 & 0.0339 & 10  & 16.3 & 1424 \\
   & Stateless  & 35.94 & 7.32 & 6.06 & 0.0278 & 80  & 16.2 & 1425 \\
\midrule
\rowcolor[gray]{0.9} 50 & Databelt   & 30.29 & 3.12 & 6.79 & 0.0330 & 0   & 17.3 & 1331 \\
   & Random     & 37.75 & 8.39 & 5.91 & 0.0265 & 30  & 16.3 & 1323 \\
   & Stateless  & 43.29 & 9.16 & 7.10 & 0.0231 & 40  & 16.2 & 1428 \\ 
\bottomrule
\end{tabular}%
}
\end{table}

\begin{figure*}[t]
    \begin{subfigure}{0.24\linewidth}
        \begin{tikzpicture}
            \begin{axis}[               
                xlabel={Input Size (MB)},
                ylabel style={yshift=-3pt,font=\footnotesize},
                xticklabel style={font=\footnotesize},  
                xlabel style={font=\footnotesize},
                yticklabel style={font=\footnotesize}, 
                ylabel={Seconds},
                xtick={10,20,30,40,50}, 
                xmin=5,
                xmax=50,
                legend style={at={(2.5,1.3)},anchor=north,draw=none,legend columns=-1},
                width=4.8cm,         
                height=4cm, 
                grid=major,
                grid style={dashed,gray!30},
                mark options={solid}
            ]
            \addplot[
                color=blue!80,
                mark=*, 
                very thick,
                mark size=1.5,
            ] coordinates {(5,5.77)(10,7.90)(15,11.02)(20,13.81)(25,16.03)(30,19.39)(35,23.03)(40,24.48)(45,27.02)(50,30.29)};
            \addlegendentry{Databelt}
            \addplot[
                color=green!60!black,
                mark=triangle*, 
                very thick,
                mark size=1.5,
                smooth
            ] coordinates {(5,7.41)(10,10.76)(15,14.89)(20,18.86)(25,21.40)(30,25.71)(35,27.53)(40,29.53)(45,34.98)(50,37.75)};
            \addlegendentry{Random}
            \addplot[
                color=orange!80,
                mark=square*, 
                very thick,
                mark size=1.5,
            ] coordinates {(5,8.26)(10,12.47)(15,16.33)(20,20.27)(25,24.06)(30,27.94)(35,32.02)(40,35.94)(45,39.42)(50,43.29)};
            \addlegendentry{Stateless}
            \end{axis}
        \end{tikzpicture}
        \caption{Latency}
        \label{fig:workflow_latency}
    \end{subfigure}
       \begin{subfigure}{0.24\linewidth}
        \begin{tikzpicture}
            \begin{axis}[               
                xlabel={Input Size (MB)},
                ylabel style={yshift=-3pt,font=\footnotesize},
                ylabel={Seconds},
                xticklabel style={font=\footnotesize},  
                xlabel style={font=\footnotesize},
                yticklabel style={font=\footnotesize}, 
                ymin=0,
                xtick={1,10,20,30,40,50}, 
                xmin=5,
                xmax=50,
                legend style={at={(1.15,1.25)},anchor=north,draw=none,legend columns=-1},
                width=4.8cm,                     
                height=4cm,
                grid=major,
                grid style={dashed,gray!30},
                mark options={solid}
            ]
            \addplot[
                color=blue!80,
                mark=*, 
                very thick,
                mark size=1.5,
            ] coordinates {(5,0.59)(10,0.64)(15,1.04)(20,1.26)(25,1.30)(30,2.04)(35,2.14)(40,2.49)(45,2.54)(50,3.12)};
            \addplot[
                color=green!60!black,
                mark=triangle*, 
                very thick,
                mark size=1.5
            ] coordinates {(5,1.30)(10,1.90)(15,2.64)(20,3.49)(25,4.04)(30,5.54)(35,5.45)(40,5.73)(45,7.54)(50,8.39)};
            \addplot[
                color=orange!80,
                mark=square*, 
                very thick,
                mark size=1.5,
            ] coordinates {(5,1.45)(10,2.43)(15,3.15)(20,4.00)(25,4.95)(30,5.67)(35,6.49)(40,7.32)(45,8.20)(50,9.16)};
            
            \end{axis}
        \end{tikzpicture}
        \caption{Read}
        \label{fig:workflow_read}
    \end{subfigure}
    \begin{subfigure}{0.24\linewidth}
        \begin{tikzpicture}
            \begin{axis}[               
                xlabel={Input Size (MB)},
                ylabel style={yshift=-3pt,font=\footnotesize},
                xticklabel style={font=\footnotesize},  
                xlabel style={font=\footnotesize},
                yticklabel style={font=\footnotesize}, 
                ylabel={Seconds},
                xtick={1,10,20,30,40,50}, 
                ymin=0,
                xmin=5,
                xmax=50,
                width=4.8cm,                    
                height=4cm,
                grid=major,
                grid style={dashed,gray!30},
                mark options={solid}
            ]
            \addplot[
                color=blue!80,
                mark=*, 
                very thick,
                mark size=1.5,
            ] coordinates {(5,1.16)(10,1.74)(15,2.41)(20,3.10)(25,3.73)(30,4.26)(35,4.76)(40,5.46)(45,6.18)(50,6.79)};
            \addplot[
                color=orange!80,
                mark=square*, 
                very thick,
                mark size=1.5,
            ] coordinates {(5,1.42)(10,2.07)(15,2.78)(20,3.43)(25,3.97)(30,4.69)(35,5.40)(40,6.06)(45,6.54)(50,7.10)};
            \addplot[
                color=green!60!black,
                mark=triangle*, 
                very thick,
                mark size=1.5,
            ] coordinates {(5,1.23)(10,1.85)(15,2.64)(20,3.31)(25,3.69)(30,3.85)(35,4.75)(40,5.13)(45,5.64)(50,5.91)};
            \end{axis}
        \end{tikzpicture}
        \caption{Write}
        \label{fig:workflow_write}
    \end{subfigure}
    \begin{subfigure}{0.24\linewidth}
        \begin{tikzpicture}
            \begin{axis}[               
                xlabel={Input Size (MB)},
                ylabel style={yshift=-6pt,font=\footnotesize},
                ylabel={Requests per second},
                xticklabel style={font=\footnotesize},  
                xlabel style={font=\footnotesize},
                yticklabel style={font=\footnotesize}, 
                xtick={10,20,30,40,50}, 
                yticklabel style={
                    /pgf/number format/fixed,
                    /pgf/number format/precision=2
                },
                xmin=5,
                xmax=50,
                width=4.8cm,                      
                height=4cm,
                grid=major,
                grid style={dashed,gray!30},
                mark options={solid}
            ]
            \addplot[
                color=blue!80,
                mark=*, 
                very thick,
                mark size=1.5,
            ] coordinates {(5,0.1733)(10,0.1266)(15,0.0907)(20,0.0724)(25,0.0624)(30,0.0516)(35,0.0434)(40,0.0409)(45,0.0370)(50,0.0330)};
            \addplot[
                color=orange!80,
                mark=square*, 
                very thick,
                mark size=1.5,
            ] coordinates {(5,0.1211)(10,0.0802)(15,0.0613)(20,0.0493)(25,0.0416)(30,0.0358)(35,0.0312)(40,0.0278)(45,0.0254)(50,0.0231)};
            \addplot[
                color=green!60!black,
                mark=triangle*, 
                thick,
            ] coordinates {(5,0.1349)(10,0.0929)(15,0.0672)(20,0.0530)(25,0.0467)(30,0.0389)(35,0.0363)(40,0.0339)(45,0.0286)(50,0.0265)};
            \end{axis}
        \end{tikzpicture}
        \caption{Throughput}
        \label{fig:workflow_throughput}
    \end{subfigure}
    \caption{Performance of serverless workflows with varying input sizes (MB), where: (a) Total workflow execution latency, (b) workflow state read latency, (c) workflow state write latency, and (d) workflow throughput for Databelt, Random, and Stateless. }
    \label{fig:propagate-t-total-performance}
\end{figure*}
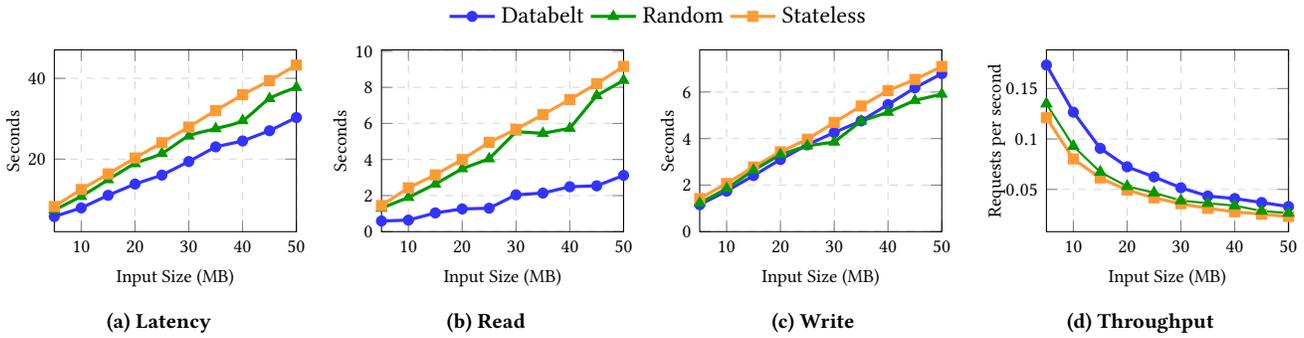
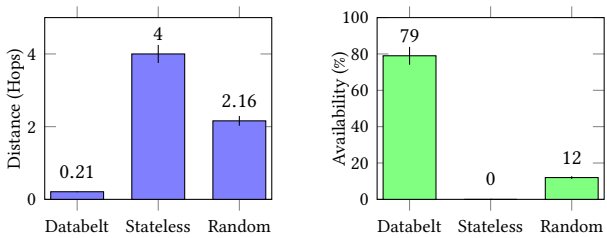
\begin{figure}[t]
    \begin{subfigure}{0.49\linewidth}
        \begin{tikzpicture}
            \begin{axis}[
                ybar,
                symbolic x coords={Databelt, Stateless, Random},
                xtick=data,
                ymin=0,
                ymax = 5,
                ylabel={Distance (Hops)},
                ylabel style={yshift=-5pt},
                xticklabel style={font=\footnotesize},  
                ylabel style={font=\footnotesize},
                yticklabel style={font=\footnotesize}, 
                bar width=20pt,
                width=4.6cm,         
                height=4cm,
                enlarge x limits=0.2,
                nodes near coords,
                nodes near coords align={vertical},
                every node near coord/.append style={font=\small, yshift=2pt}
            ]
                \addplot[fill=blue!50, error bars/.cd, y dir=both,
                    y explicit, error mark=none ] coordinates {
                    (Databelt,0.21) +- (0,0.0130)
                    (Stateless,4) +- (0,0.248)
                    (Random,2.16) +- (0,0.134)
                };
            \end{axis}
        \end{tikzpicture}
        \label{fig:distance_hops}
    \end{subfigure}
    \hfill
    \begin{subfigure}{0.49\linewidth}
        \begin{tikzpicture}
            \begin{axis}[
                area legend,
                ybar,
                symbolic x coords={Databelt, Stateless, Random},
                xtick=data,
                ymin=0,
                ymax = 100,
                ylabel={Availability (\%)},
                ylabel style={yshift=-9pt},
                xlabel style={font=\small},
                xticklabel style={font=\footnotesize},  
                ylabel style={font=\footnotesize},
                yticklabel style={font=\footnotesize}, 
                bar width=20pt,
                width=4.6cm,         
                height=4cm,
                enlarge x limits=0.2,
                nodes near coords,
                nodes near coords align={vertical},
                every node near coord/.append style={font=\small,yshift=2pt} 
            ]
                \addplot[fill=green!50,
                error bars/.cd,
                y dir=both,
                y explicit,
                error mark=none
                ] coordinates {
                (Databelt,79) +- (0,4.9)
                (Stateless,0) +- (0,0.0)
                (Random,12) +- (0,0.74)
                };
            \end{axis}
        \end{tikzpicture}
        \label{fig:availability}
    \end{subfigure}
    \caption{State propagation performance for Databelt, Stateless, and Random, where: (a) the mean state read distance in network hops, and (b) local state availability.}
    \label{fig:propagate-local-state-availability-bar}
\end{figure}

\begin{figure}
    \centering
    \begin{tikzpicture}
        \begin{axis}[
            width=6cm,
            height=4.5cm,
            xticklabel style={font=\footnotesize},  
            ylabel style={font=\footnotesize},
            yticklabel style={font=\footnotesize}, 
            colormap/viridis,
            colorbar,
            point meta min=0,
            point meta max=100,
            ylabel={Input Size (MB)},
            xtick={0,1,2},
            xticklabels={Databelt, Random, Stateless},
            ytick={0,...,5},                 
            yticklabels={1,10,20,30,40,50},  
            enlargelimits=false,
            nodes near coords,
            nodes near coords align={center},
            every node near coord/.append style={
                font=\footnotesize,
                text=white
            }
        ]
        \addplot [
            matrix plot,
            mesh/cols=3,
            mesh/rows=6,
            point meta=explicit
        ] coordinates {
            (0,0) [0]   (1,0) [30]   (2,0) [100]
            (0,1) [0]   (1,1) [100]  (2,1) [100]
            (0,2) [0]   (1,2) [40]   (2,2) [80]
            (0,3) [0]   (1,3) [30]   (2,3) [100]
            (0,4) [0]   (1,4) [10]   (2,4) [80]
            (0,5) [0]   (1,5) [30]   (2,5) [40]
        };

        \addplot [
            only marks,
            nodes near coords,
            point meta=explicit,
            every node near coord/.append style={
                font=\footnotesize\bfseries,
                text=black
            }
        ] coordinates {
            (1,1) [100] (2,0) [100] (2,1) [100] (2,3) [100]
        };

        \end{axis}
    \end{tikzpicture}
    \caption{SLO violations (\%) for state propagation with defined SLO of 60 ms for different input sizes (MB) for Databelt, Random, and Stateless.}
    \label{fig:propagate-performance-slo}
\end{figure}
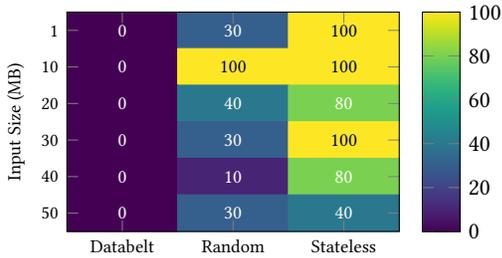

\begin{figure}[t]
    \begin{subfigure}{0.49\linewidth}
        \begin{tikzpicture}
            \begin{axis}[               
                xlabel={Input Size (MB)},
                ylabel style={yshift=-3pt,font=\footnotesize},
                xticklabel style={font=\footnotesize},  
                yticklabel style={font=\footnotesize},  
                ylabel style={font=\footnotesize},
                xlabel style={font=\footnotesize},
                ylabel={CPU Usage (\%)},
                xtick={0,10,20,30,40,50}, 
                legend style={at={(1.2,1.35)},anchor=north,draw=none,legend columns=-1},
                xmin=0,
                xmax=50,
                ymin=12,
                ymax=18,
                width=4.5cm,         
                height=4cm,
                grid=major,
                grid style={dashed,gray!30},
                mark options={solid}
            ]
            \addplot[
                color=blue!80,
                very thick,
            ] coordinates {(0,12.7)(5,17.2)(10,16.4)(15,17.0)(20,17.0)(25,17.1)(30,17.3)(35,16.9)(40,17.2)(45,17.1)(50,17.3)};
            \addlegendentry{Databelt}
            \addplot[
                color=green!60!black,
                very thick,
            ] coordinates {(50,16.3)(45,16.3)(40,16.3)(35,16.3)(30,16.3)(25,16.3)(20,16.4)(15,16.4)(10,16.4)(5,16.3)(0,13.6)};
            \addlegendentry{Random}
            \addplot[
                color=orange!80,
                very thick,
            ] coordinates {(50,16.2)(45,16.2)(40,16.2)(35,16.2)(30,16.2)(25,16.3)(20,16.3)(15,16.3)(10,16.5)(5,16.3)(0,13.6)};
            \addlegendentry{Stateless}
            \end{axis}
        \end{tikzpicture}
        \caption{CPU}
    \end{subfigure}
    \hfill
    \begin{subfigure}{0.49\linewidth}
        \begin{tikzpicture}
            \begin{axis}[               
                xlabel={Input Size (MB)},
                ylabel style={yshift=-3pt},
                xticklabel style={font=\footnotesize},  
                yticklabel style={font=\footnotesize},  
                ylabel style={font=\footnotesize},
                xlabel style={font=\footnotesize}, 
                ylabel={RAM (GB)},
                xtick={0,10,20,30,40,50},
                xmin=0,
                xmax=50,
                ymin=1.15,
                ymax=1.45,
                width=4.5cm,         
                height=4cm,
                grid=major,
                grid style={dashed,gray!30},
                mark options={solid}
            ]
            \addplot[
                color=blue!80,
                very thick,
            ] coordinates {(0,1.197)(5,1.285)(10,1.289)(15,1.296)(20,1.293)(25,1.300)(30,1.302)(35,1.303)(40,1.308)(45,1.314)(50,1.300)};
            
            \addplot[
                color=orange!80,
                very thick,
            ] coordinates {
                (0,1.291)
                (5,1.391)
                (10,1.391)
                (15,1.391)
                (20,1.391)
                (25,1.391)
                (30,1.392)
                (35,1.394)
                (40,1.395)
                (45,1.395)
                (50,1.397)
            };
            
            \addplot[
                color=green!60!black,
                very thick,
            ] coordinates {
                (0,1.292)
                (5,1.392)
                (10,1.392)
                (15,1.392)
                (20,1.392)
                (25,1.391)
                (30,1.392)
                (35,1.392)
                (40,1.393)
                (45,1.394)
                (50,1.292)
            };

            \end{axis}
        \end{tikzpicture}
        \caption{RAM}
    \end{subfigure}
    \caption{Resource usage during the state propagation performance experiment for Databelt, Random, and Stateless, where: (a) CPU, and (b) RAM.}
    \label{fig:propagate-performance-resources}
\end{figure}
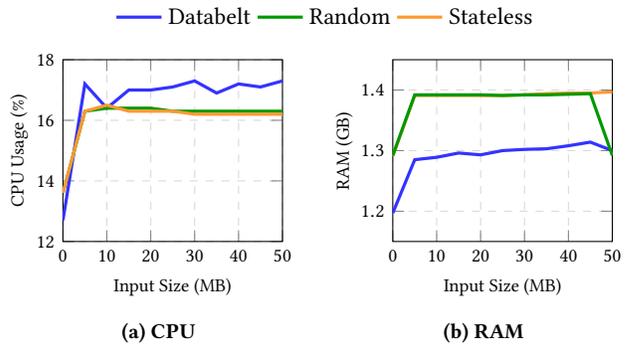

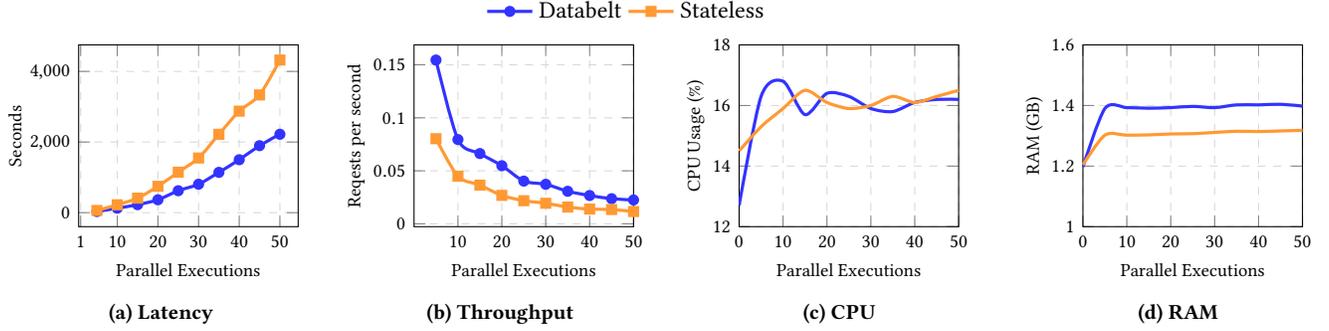
\begin{figure*}[h]
    \begin{subfigure}{0.24\linewidth}
        \begin{tikzpicture}
            \begin{axis}[               
                xlabel={Parallel Executions},
                ylabel={Seconds},
                ylabel style={yshift=-3pt,font=\footnotesize},
                xticklabel style={font=\footnotesize}, 
                yticklabel style={font=\footnotesize}, 
                xlabel style={font=\footnotesize},
                xtick={1,10,20,30,40,50},
                legend style={at={(2.5,1.3)},anchor=north,draw=none,legend columns=-1},
                width=4.5cm,         
                height=4cm,
                grid=major,
                grid style={dashed,gray!30},
                mark options={solid}
            ]
            \addplot[
                color=blue!80,
                mark=*, 
                very thick,
                mark size=1.5,
                smooth
            ] coordinates {(5,32)(10,126)(15,226)(20,365)(25,621)(30,805)(35,1141)(40,1499)(45,1897)(50,2220)};
            \addlegendentry{Databelt}
            \addplot[
                color=orange!80,
                mark=square*, 
                very thick,
                mark size=1.5,
                smooth
            ] coordinates {(5,62)(10,223)(15,411)(20,746)(25,1146)(30,1548)(35,2219)(40,2874)(45,3334)(50,4322)};
            \addlegendentry{Stateless}
            \end{axis}
        \end{tikzpicture}
        \caption{Latency}
        \label{fig:scalability_latency}
    \end{subfigure}
    \hfill
    \begin{subfigure}{0.24\linewidth}
        \begin{tikzpicture}
            \begin{axis}[               
                xlabel={Parallel Executions},
                ylabel style={yshift=-3pt,font=\footnotesize},
                ylabel={Reqests per second},
                xticklabel style={font=\footnotesize},  
                yticklabel style={font=\footnotesize}, 
                xlabel style={font=\footnotesize},
                xtick={10,20,30,40,50}, 
                xmin=0,
                xmax=50,
                yticklabel style={
                    /pgf/number format/fixed,
                    /pgf/number format/precision=2
                },
                width=4.5cm,
                height=4cm,
                grid=major,
                grid style={dashed,gray!30},
                mark options={solid}
            ]
            \addplot[
                color=blue!80,
                mark=*, 
                very thick,
                mark size=1.5,
                smooth
            ] coordinates {(5,0.1545)(10,0.0795)(15,0.0663)(20,0.0548)(25,0.0403)(30,0.0373)(35,0.0307)(40,0.0267)(45,0.0237)(50,0.0225)};
            \addplot[
                color=orange!80,
                mark=square*, 
                very thick,
                mark size=1.5,
                smooth
            ] coordinates {(5,0.0803)(10,0.0449)(15,0.0365)(20,0.0268)(25,0.0218)(30,0.0194)(35,0.0158)(40,0.0139)(45,0.0135)(50,0.0116)};
            \end{axis}
        \end{tikzpicture}
        \caption{Throughput}
        \label{fig:scalability_throughput}
    \end{subfigure}
    \hfill
    \begin{subfigure}{0.24\linewidth}
        \begin{tikzpicture}
            \begin{axis}[               
                xlabel={Parallel Executions},
                ylabel style={yshift=-3pt,font=\footnotesize},
                ylabel={CPU Usage (\%)},
                xtick={0,10,20,30,40,50}, 
                xticklabel style={font=\footnotesize},  
                xlabel style={font=\footnotesize},
                yticklabel style={font=\footnotesize}, 
                legend style={at={(1.2,1.25)},anchor=north,draw=none,legend columns=-1},
                xmin=0,
                xmax=50,
                ymin=12,
                ymax=18,
                width=4.5cm,
                height=4cm,
                grid=major,
                grid style={dashed,gray!30},
                mark options={solid}
            ]
            \addplot[
                color=blue!80,
                very thick,
                smooth
            ] coordinates {(50,16.2)(45,16.2)(40,16.1)(35,15.8)(30,15.9)(25,16.3)(20,16.4)(15,15.7)(10,16.8)(5,16.3)(0,12.7)};
            \addplot[
                color=orange!80,
                very thick,
                smooth
            ] coordinates {(0,14.5)(5,15.3)(10,15.9)(15,16.5)(20,16.1)(25,15.9)(30,16.0)(35,16.3)(40,16.1)(45,16.3)(50,16.5)};
            \end{axis}
        \end{tikzpicture}
        \caption{CPU}
        \label{fig:scalability_cpu}
    \end{subfigure}
    \hfill
    \begin{subfigure}{0.24\linewidth}
        \begin{tikzpicture}
            \begin{axis}[               
                xlabel={Parallel Executions},
                ylabel style={yshift=-3pt,font=\footnotesize},
                xticklabel style={font=\footnotesize},  
                ylabel style={font=\footnotesize},
                xlabel style={font=\footnotesize},
                yticklabel style={font=\footnotesize}, 
                ylabel={RAM (GB)},
                xtick={0,10,20,30,40,50},
                xmin=0,
                xmax=50,
                ymin=1,
                ymax=1.6,
                width=4.5cm,
                height=4cm,
                grid=major,
                grid style={dashed,gray!30},
                mark options={solid}
            ]
            \addplot[
                color=blue!80, 
                very thick,
                smooth
            ] coordinates {
                (50,1.398) (45,1.404) (40,1.402) (35,1.402) (30,1.393) 
                (25,1.397) (20,1.393) (15,1.391) (10,1.393) (5,1.390) (0,1.197)
            };
            
            \addplot[
                color=orange!80,
                very thick,
                smooth
            ] coordinates {
                (0,1.204) (5,1.302) (10,1.302) (15,1.303) (20,1.306) 
                (25,1.307) (30,1.311) (35,1.315) (40,1.314) (45,1.316) (50,1.318)
            };
            \end{axis}
        \end{tikzpicture}
        \caption{RAM}
        \label{fig:scalability_ram}
    \end{subfigure}
    \caption{Scalability Results of Databelt with increasing parallel workflow executions, compared to Stateless approaches. Subfigures show (a) Total workflow latency, (b) Total workflow throughput, (c) CPU usage, and (d) RAM usage.}
    \label{fig:propagate-t-total-scalability}
\end{figure*}

\subsection{Function State Propagation Performance Results}

\paragraph{Workflow Execution Time}
\color{black}{Overall, Databelt improves the workflow execution time by up to 66\% and increases the throughput by up to 50\%.}
~\cref{fig:propagate-t-total-performance} presents the latency results of the performance experiments for serverless workflows with varying state input sizes. \cref{tab:function_state_results} shows detailed results from the function state propagation experiments with function state fusion depth 1. 
~\cref{fig:workflow_latency} shows the workflow execution time on the $y$-axis and the state size on the $x$-axis.  Databelt decreases the total workflow latency by 33\% compared to Stateless and by 22\% compared to Random storage placement strategies. 
~\cref{fig:workflow_read} shows that Databelt reduces the state read time compared to Stateless by 66\% and compared to Random by 62\%. 
In ~\cref{fig:workflow_write}, Databelt write time is 9\% lower than Stateless, and 4\% higher than Random. Databelt significantly reduces state read times while performing similarly regarding state write time, due to the computational overhead incurred for the state propagation mechanisms. 
~\cref{fig:workflow_throughput} reports the throughput. Databelt has consistently higher throughput than the baselines. Databelt improves throughput by 50\% compared to Stateless and 29\% compared to Random.

\paragraph{State Availability}
\color{black}{Databelt presents a local state availability of 79\% and reduces the hops distance to an average of 0.21.}
\cref{fig:propagate-local-state-availability-bar} shows the mean state proximity to the function and local state availability.
\cref{fig:distance_hops} presents the state distance determined by the number of nodes between the function and the state. Databelt shows a lower mean distance (0.21~hops) to retrieve the state compared to Stateless (4~hops) and Random (2.16~hops), respectively. \cref{fig:availability} shows the state availability. The state is available if the function can access it locally once the function executes. Stateless retrieves the data from cloud nodes, while random leverages local storage on neighbor satellite nodes; therefore, the average distance is lower than the stateless approach. Databelt has a local state availability of 79\% compared to 12\% with the Random.

\paragraph{SLO Compliance}
\color{black}{Databelt demonstrates zero SLO violations of the defined 60~ms across all input sizes, while Random and Stateless show significant violations.}
\cref{fig:propagate-performance-slo} depicts the SLO violations between Databelt and the baselines. Databelt respects the SLO constraints, while Random has a high variation in SLO violations from 10\% to 100\%, explained by the nature of picking random nodes in the network path. Stateless has many SLO violations due to the high latency to read and write from the satellite to the storage, typically located in cloud services.

\paragraph{Resource Usage}
\color{black}{Databelt slightly reduces the memory usage by 7\%.}
\cref{fig:propagate-performance-resources} shows the average resource usage over all nodes during experiment execution. Although Databelt shows a slight CPU increase compared to the baselines, both CPU usage and RAM allocation remain linear without peaks. The additional computational overhead incurred by Databelt is a tradeoff to provide the state to the function.

\subsection{State Propagation Scalability Results}

\paragraph{Parallel Executions}
\color{black}{Databelt reduces the latency by 47\% and increases the throughput by up to 91\%.}
~\cref{fig:propagate-t-total-scalability} shows the scalability results for the state propagation experiment with a fixed state size of 2MB, where the $x$-axis represents the fan-out degree and the $y$-axis represents the total workflow latency in seconds. In \cref{fig:scalability_latency}, we see Databelt reduces workflow latency by 47\% compared to Stateless. ~\cref{fig:scalability_throughput} shows the throughput, where Databelt consistently achieves higher throughput than Stateless. Databelt increases throughput by up to 91\%.

\paragraph{Resource Usage}
~\cref{fig:scalability_cpu} and \cref{fig:scalability_ram} show CPU and RAM usage during parallel execution experiments, respectively. Throughout the experiment, both CPU usage and RAM allocation remain stable, showing a slight CPU usage increase and a slight RAM usage decrease.

\cref{tab:state_propagation_scalability} shows the detailed results for parallel executions and resource usage of the state fusion scalability experiments with function fusion depth 1.

\subsection{Function State Fusion Results}

\paragraph{Function Fusion Degree}
Databelt reduces latency by 20\% compared to the Baseline for stateless and 19\% for stateful functions.
\cref{fig:bundle-size-tt-bar} shows the total workflow execution time with varying function fusion degrees, where Databelt groups a specific number of functions in one single sandbox, and the Baseline retrieves the data for every single function with a fixed input size of 10MB. 
In \cref{fig:bundle_stateless}, functions are stateless, which means they need to retrieve the state from remote storage, while in \cref{fig:bundle_stateful}, functions are stateful (i.e., function states are locally available).
Due to the growing amount of requests and, consequently, growing overhead, we observe that $|f|$ grows proportionally with the function latency between Databelt and the Baseline. 

\begin{table}[t]
\caption{Scalability results of state propagation with state fusion depth 1.}
\label{tab:state_propagation_scalability}
\resizebox{\columnwidth}{!}{%
\begin{tabular}{cccccc}
\toprule
\textbf{Parallel Executions} & \textbf{System} & \textbf{Latency (sec)} & \textbf{RPS} & \textbf{CPU (\%)} & \textbf{RAM (MB)} \\
\midrule
\rowcolor[gray]{0.9} 5  & Databelt   & 32   & 0.1545 & 16.3 & 1390 \\
   & Stateless  & 62   & 0.0803 & 15.3 & 1302 \\
\midrule
\rowcolor[gray]{0.9} 10 & Databelt   & 126  & 0.0795 & 16.8 & 1393 \\
   & Stateless  & 223  & 0.0449 & 15.9 & 1302 \\
\midrule
\rowcolor[gray]{0.9} 15 & Databelt   & 226  & 0.0663 & 15.7 & 1391 \\
   & Stateless  & 411  & 0.0365 & 16.5 & 1303 \\
\midrule
\rowcolor[gray]{0.9} 20 & Databelt   & 365  & 0.0548 & 16.4 & 1393 \\
   & Stateless  & 746  & 0.0268 & 16.1 & 1306 \\
\midrule
\rowcolor[gray]{0.9} 25 & Databelt   & 621  & 0.0403 & 16.3 & 1397 \\
   & Stateless  & 1146 & 0.0218 & 15.9 & 1307 \\
\midrule
\rowcolor[gray]{0.9} 30 & Databelt   & 805  & 0.0373 & 15.9 & 1393 \\
   & Stateless  & 1548 & 0.0194 & 16.0 & 1311 \\
\midrule
\rowcolor[gray]{0.9} 35 & Databelt   & 1141 & 0.0307 & 15.8 & 1402 \\
   & Stateless  & 2219 & 0.0158 & 16.3 & 1315 \\
\midrule
\rowcolor[gray]{0.9} 40 & Databelt   & 1499 & 0.0267 & 16.1 & 1402 \\
   & Stateless  & 2874 & 0.0139 & 16.1 & 1314 \\
\midrule
\rowcolor[gray]{0.9} 45 & Databelt   & 1897 & 0.0237 & 16.2 & 1404 \\
   & Stateless  & 3334 & 0.0135 & 16.3 & 1316 \\
\midrule
\rowcolor[gray]{0.9} 50 & Databelt   & 2220 & 0.0225 & 16.2 & 1398 \\
   & Stateless  & 4322 & 0.0116 & 16.5 & 1318 \\
\bottomrule
\end{tabular}%
}
\end{table}

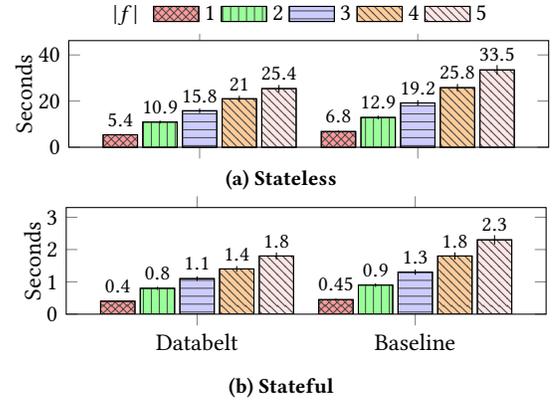
\begin{figure}[t]
    \begin{subfigure}{\linewidth}
        \centering
        \begin{tikzpicture}
            \begin{axis}[
                ybar,
                symbolic x coords={Bundled, Single},
                x=2.9cm,
                enlarge x limits=0.6,
                xtick=data,
                ymin=0,
                ylabel={Seconds},
                ylabel style={yshift=-4pt},
                bar width=13pt,
                xticklabels={},
                width=7cm,
                ymax=42,
                height=3cm,  
                area legend,
                legend style={at={(0.9,1.25)},anchor=east,draw=none,legend columns=-1},
                enlarge y limits={upper, value=0.1},
                nodes near coords,
                nodes near coords align={vertical},
                every node near coord/.append style={font=\small} 
            ]
                \addlegendimage{empty legend},
                \addlegendentry{\hspace{-.7cm}\small{$|f|$}}
                \addplot[ybar, fill=red!40, postaction={pattern=crosshatch,pattern color=black!70},error bars/.cd, y dir=both, y explicit, error mark=none] coordinates {
                (Single,6.8) +-(0,0.422)
                (Bundled,5.4) +-(0,0.335)
                };
                \addlegendentry{1}
                
                \addplot[ybar, fill=green!40, postaction={pattern=vertical lines,pattern color=black!70},error bars/.cd, y dir=both, y explicit, error mark=none] coordinates {(Single,12.9)+-(0,0.800) (Bundled,10.9)+-(0,0.676)};
                \addlegendentry{2}
                \addplot[ybar, fill=blue!20, postaction={pattern=horizontal lines,pattern color=black!70},error bars/.cd, y dir=both, y explicit, error mark=none] coordinates {(Single,19.2) +-(0,1.190) (Bundled,15.8)+-(0,0.980)};
                \addlegendentry{3}
                \addplot[ybar, fill=orange!40, postaction={pattern=north west lines,pattern color=black!70},error bars/.cd, y dir=both, y explicit, error mark=none] coordinates {(Single,25.8)+-(0,1.600)(Bundled,21)+-(0,1.302)};
                \addlegendentry{4}
                \addplot[ybar, fill=pink!40, postaction={pattern=north west lines,pattern color=black!70},error bars/.cd, y dir=both, y explicit, error mark=none] coordinates {(Single,33.5)+-(0,2.077)(Bundled,25.4)+-(0,1.575)};
                \addlegendentry{5}
            \end{axis}
        \end{tikzpicture}
        \vspace{-1em}
        \caption{Stateless}
        \label{fig:bundle_stateless}
    \end{subfigure}
    \begin{subfigure}{\linewidth}
        \centering
        \begin{tikzpicture}
            \begin{axis}[
                ybar,
                symbolic x coords={Databelt, Baseline},
                xtick=data,
                x=2.9cm,
                enlarge x limits=0.6,
                ymin=0,
                ymax=3,
                ylabel={Seconds},
                ylabel style={yshift=-4pt},
                ytick={0,1,2,3},
                width=7cm,
                height=3cm,  
                bar width=13pt,
                enlarge y limits={upper, value=0.1},
                nodes near coords,
                nodes near coords align={vertical},
                every node near coord/.append style={font=\small} 
            ]
               
                \addplot[ybar, fill=red!40, postaction={pattern=crosshatch,pattern color=black!70},error bars/.cd, y dir=both, y explicit, error mark=none] coordinates {(Baseline,0.45)+-(0,0.028) (Databelt,0.4)+-(0,0.025)};
                \addplot[ybar, fill=green!40, postaction={pattern=vertical lines,pattern color=black!70},error bars/.cd, y dir=both, y explicit, error mark=none] coordinates {(Baseline,0.9)+-(0,0.056) (Databelt,0.8)+-(0,0.050)};
                \addplot[ybar, fill=blue!20, postaction={pattern=horizontal lines,pattern color=black!70},error bars/.cd, y dir=both, y explicit, error mark=none] coordinates {(Baseline,1.3)+-(0,0.081) (Databelt,1.1)+-(0,0.068)};
                \addplot[ybar, fill=orange!40, postaction={pattern=north west lines,pattern color=black!70},error bars/.cd, y dir=both, y explicit, error mark=none] coordinates {(Baseline,1.8)+-(0,0.112) (Databelt,1.4)+-(0,0.087)};
                \addplot[ybar, fill=pink!40, postaction={pattern=north west lines,pattern color=black!70},error bars/.cd, y dir=both, y explicit, error mark=none] coordinates {(Baseline,2.3)+-(0,0.143) (Databelt,1.8)+-(0,0.112)};
            \end{axis}
        \end{tikzpicture}
        \caption{Stateful}
        \label{fig:bundle_stateful}
    \end{subfigure}
    \caption{Total function latency for varying function fusion depths in Databelt compared to the Baseline, where every function retrieves its single state. Subfigures show (a) Stateless latency, where functions retrieve the state from remote storage, and (b) stateful latency to retrieve the data locally.}
    \label{fig:bundle-size-tt-bar}
\end{figure}
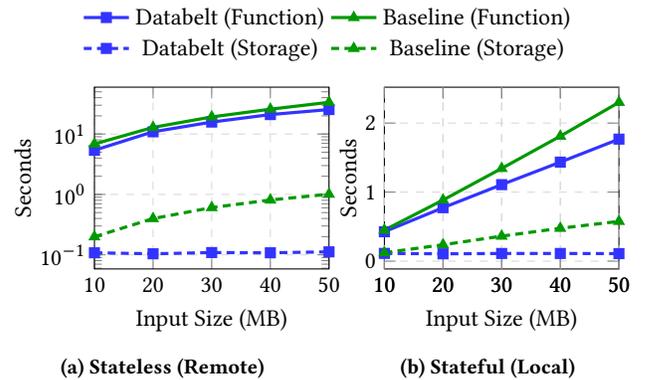
\begin{figure}[h!]
    \begin{subfigure}{0.49\linewidth}
        \begin{tikzpicture}
            \begin{axis}[               
                xlabel={Input Size (MB)},
                ylabel style={yshift=-3pt},
                ylabel={Seconds},
                xtick={10,20,30,40,50}, 
                xticklabels={10, 20, 30, 40, 50}, 
                xmin=10,
                xmax=50,
                legend style={at={(1,1.5)},anchor=north,draw=none,legend columns=2},
                width=4.7cm,         
                height=4cm,
                ymode=log,
                grid=major,
                grid style={dashed,gray!30},
                mark options={solid}
            ]

            \addplot[
                color=blue!80,
                mark=square*, 
                very thick,
                mark size=1.5,
            ] coordinates {(10,5.379)(20,10.889)(30,15.776)(40,20.997)(50,25.417)};
            \addlegendentry{Databelt (Function)}
            \addplot[
                color=green!60!black,
                mark=triangle*, 
                very thick,
                mark size=1.5,
            ] coordinates {(10,6.839)(20,12.876)(30,19.199)(40,25.767)(50,33.507)};
            \addlegendentry{Baseline (Function)}

            \addplot[
                color=blue!80,
                mark=square*, 
                very thick,
                mark size=1.5,
                dash pattern=on 3pt off 2pt,
            ] coordinates {(10,0.108)(20,0.104)(30,0.109)(40,0.108)(50,0.112)};
            \addlegendentry{Databelt (Storage)}
            \addplot[
                color=green!60!black,
                mark=triangle*, 
                very thick,
                mark size=1.5,
                dash pattern=on 3pt off 2pt,
            ] coordinates {(10,0.199)(20,0.398)(30,0.605)(40,0.809)(50,1.008)};
            \addlegendentry{Baseline (Storage)}
            \end{axis}
        \end{tikzpicture}
        \caption{Stateless (Remote)}
        \label{fig:bundled_latency_stateless}
    \end{subfigure}
    \hfill
    \begin{subfigure}{0.49\linewidth}
        \begin{tikzpicture}
            \begin{axis}[               
                xlabel={Input Size (MB)},
                ylabel style={yshift=-3pt},
                ylabel={Seconds},
                xtick={10,20,30,40,50}, 
                extra x ticks={10,20,30,40,50}, 
                xmin=10,
                xmax=50,
                width=4.7cm,         
                height=4cm,
                grid=major,
                grid style={dashed,gray!30},
                mark options={solid}
            ]
            \addplot[
                color=blue!80,
                mark=square*, 
                very thick,
                mark size=1.5,
            ] coordinates {(10,0.426)(20,0.774)(30,1.108)(40,1.434)(50,1.767)};
            \addplot[
                color=green!60!black,
                mark=triangle*, 
                very thick,
                mark size=1.5,
            ] coordinates {(10,0.449)(20,0.885)(30,1.343)(40,1.810)(50,2.300)};

            \addplot[
                color=blue!80,
                mark=square*, 
                very thick,
                mark size=1.5,
                dash pattern=on 3pt off 2pt
            ] coordinates {(10,0.112)(20,0.106)(30,0.111)(40,0.111)(50,0.109)};
            \addplot[
                color=green!60!black,
                mark=triangle*, 
                very thick,
                mark size=1.5,
                dash pattern=on 3pt off 2pt
            ] coordinates {(10,0.122)(20,0.237)(30,0.362)(40,0.476)(50,0.576)};
            \end{axis}
        \end{tikzpicture}
        \caption{Stateful (Local)}
        \label{fig:bundled_latency_stateful}
    \end{subfigure}
    \caption{Total function latency comparison between Databelt and the Baseline. Subfigures illustrate (a) stateless latency with remote storage access and (b) stateful latency with local storage access.}
    \label{fig:databelt-t-toal-performance}
\end{figure}

\begin{table}[t]
\caption{Function state function results with latency per–depth function and storage for stateless (remote storage) and stateful (local storage).}
\label{tab:state_fusion_results_table}
\centering
\resizebox{\columnwidth}{!}{%
\begin{tabular}{cclccc}
\toprule
\textbf{Depth} & \textbf{Input Size (MB)} & \textbf{System} & \textbf{State} & \textbf{Function (s)} & \textbf{Storage (s)} \\
\midrule
\rowcolor[gray]{0.9} 1 & 10 & Databelt & Stateless & 5.379 & 0.108 \\
1 & 10 & Remote Storage & Stateless & 6.839 & 0.199 \\
\rowcolor[gray]{0.9} 1 & 10 & Databelt & Stateful  & 0.426 & 0.112 \\
1 & 10 & Local Storage  & Stateful  & 0.449 & 0.122 \\
\midrule
\rowcolor[gray]{0.9} 2 & 20 & Databelt & Stateless & 10.889 & 0.104 \\
2 & 20 & Remote Storage & Stateless & 12.876 & 0.398 \\
\rowcolor[gray]{0.9} 2 & 20 & Databelt & Stateful  & 0.774 & 0.106 \\
2 & 20 & Local Storage  & Stateful  & 0.885 & 0.237 \\
\midrule
\rowcolor[gray]{0.9} 3 & 30 & Databelt & Stateless & 15.776 & 0.109 \\
3 & 30 & Remote Storage & Stateless & 19.199 & 0.605 \\
\rowcolor[gray]{0.9} 3 & 30 & Databelt & Stateful  & 1.108 & 0.111 \\
3 & 30 & Local Storage  & Stateful  & 1.343 & 0.362 \\
\midrule
\rowcolor[gray]{0.9} 4 & 40 & Databelt & Stateless & 20.997 & 0.108 \\
4 & 40 & Remote Storage & Stateless & 25.767 & 0.809 \\
\rowcolor[gray]{0.9} 4 & 40 & Databelt & Stateful  & 1.434 & 0.111 \\
4 & 40 & Local Storage  & Stateful  & 1.810 & 0.476 \\
\midrule
\rowcolor[gray]{0.9} 5 & 50 & Databelt & Stateless & 25.417 & 0.112 \\
5 & 50 & Remote Storage & Stateless & 33.507 & 1.008 \\
\rowcolor[gray]{0.9} 5 & 50 & Databelt & Stateful  & 1.767 & 0.109 \\
5 & 50 & Local Storage  & Stateful  & 2.300 & 0.576 \\
\bottomrule
\end{tabular}%
}
\end{table}

\paragraph{Function and Storage Access Latency}
Databelt minimizes the storage access latency, maintaining a constant overhead, regardless of the function fusion depth.
~\cref{fig:databelt-t-toal-performance} presents results with varying input state sizes up to 50MB. The latency is shown on the $y$-axis and the state input size on the $x$-axis. Databelt shows lower latency for the complete function processing (solid lines)  with performance increase according to the state input size on the $x$-axis for both stateless (\cref{fig:bundled_latency_stateless}) and stateful (\cref{fig:bundled_latency_stateful}). In \cref{fig:databelt-t-toal-performance}, we isolate the storage access latency (traced lines) for Databelt and Baseline. Due to the additional storage read and write operations for the Baseline functions, the storage access latency increases proportionally as the function input size increases for both stateless (\cref{fig:bundled_latency_stateless}) and stateful (\cref{fig:bundled_latency_stateful}). On the other hand, the Databelt fusion mechanism results in a constant overhead since the number of storage operations stays the same regardless of the function depth.
\cref{tab:state_fusion_results_table} shows the detailed results for the function state fusion with function and storage execution without state propagation.

\subsection{Databelt Service Scalability Results}

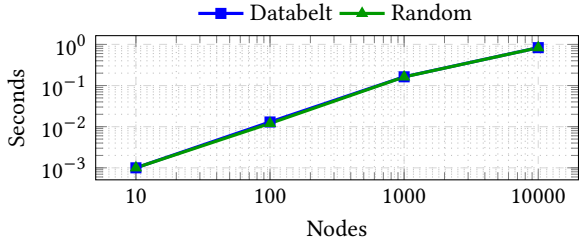
\begin{figure}[t]
    \begin{tikzpicture}
        \begin{axis}[
            xlabel={Nodes},
            ylabel={Seconds},
            width=8cm,
            height=3.5cm,
            grid=both,
            grid style={dashed,gray!30},
            minor grid style={dotted,gray!50},
            legend style={at={(0.5,1.3)},anchor=north,draw=none,legend columns=-1},
            mark options={solid},
            xmode=log,
            ymode=log,
            xtick=data,
            xticklabels={$10$, $100$, 1000, 10000},
            minor x tick num=9
        ]
            \addplot[
                blue,
                mark=square*, 
                very thick,
                mark size=1.5
            ] coordinates {
                (10,0.001) (100,0.013) (1000,0.163) (10000,0.833)
            };
            \addlegendentry{Databelt}
            \addplot[
                color=green!60!black,
                mark=triangle*, 
                very thick,
                mark size=1.5
            ] coordinates {
                (10,0.001) (100,0.012) (1000,0.162) (10000,0.830)
            };
            \addlegendentry{Random}
        \end{axis}
    \end{tikzpicture}
    \caption{Simulation of Databelt state propagation storage node election from 10 to 10,000 nodes.}
    \label{fig:databelt-scalability}
\end{figure}

\paragraph{Databelt Latency}
Due to the complex and high costs of executing multiple nodes in the 3D continuum, for the scalability experiments, we evaluate Databelt by simulating thousands of heterogeneous nodes and executing the simulation on the cloud-type node as described in \cref{tab:testbed-config}. 
Databelt maintains efficient policy runtimes with minimal overhead, even as network topology scales to 10,000 nodes.
~\cref{fig:databelt-scalability} illustrates the runtime performance of Databelt state migration across different network sizes. Even as the network expands, Databelt maintains a performance level similar to the Random state placement because it selects a subset of eligible candidate nodes, which reduces computational overhead.

\subsection{Discussion}

\paragraph{Orbital Dynamics and Network Modeling} 
Databelt relies on a system-wide topology view that reflects the current state of the 3D Compute Continuum. While this view enables real-time state propagation decisions, it assumes up-to-date network information is available. In our evaluation, we use a physical testbed composed of eight ARM-based nodes connected via a local network with controlled latency injection using Linux \texttt{tc}, allowing us to emulate dynamic network conditions. However, accurately modeling orbital dynamics would require an emulator that supports i) the simulation of satellite movement, ii) the execution of application-layer workloads, and iii) the simulation of Edge nodes. While there are satellite emulators that support the first requirement, e.g., Hypatia~\cite{Hypatia2020}, requirements one and three, e.g., Stardust~\cite{stardust2025}, or the first two requirements, e.g., Celestial~\cite{Celestial2022}, StarryNet~\cite{StarryNet2023}, to the best of our knowledge, there is currently no emulator that supports all three. Consequently, we approximate orbital dynamics by varying network latency and node reachability at runtime.  While this approach allows us to evaluate Databelt’s behavior under network dynamic conditions, we acknowledge that it does not capture the full complexity of orbital mechanics. As such, the absence of orbital modeling may limit the fidelity of the evaluation when compared to large-scale LEO constellations in operation. Nevertheless, our experimental setup mimics satellite movement and connectivity, and provides a basis for evaluating Databelt’s mechanisms in dynamic network environments.

\paragraph{Scalability}
To enable high scalability in large-scale networks such as the 3D Continuum, Databelt explicitly separates concerns across its three-phase propagation model: the \emph{Identify} and \emph{Compute} phases operate entirely in the control plane and compute state propagation decisions asynchronously, while the \emph{Offload} phase executes these decisions at runtime via lightweight API calls. This separation ensures that the function execution path remains unaffected by control-plane operations. 
Nevertheless, our evaluation simulates up to 10,000 nodes and multiple concurrent workflows on a centralized control-plane setup. While this provides a sound approximation of Databelt’s scalability under load, distributed control-plane deployments or extreme-scale topologies may introduce coordination overheads not captured in our current setup. While Databelt is designed to scale across large and heterogeneous topologies, real-world deployments may exhibit different scalability characteristics depending on the underlying infrastructure and control-plane distribution.

\paragraph{Security and Fault Tolerance}
Databelt relies on the underlying serverless platform mechanisms, such as encryption, access control, and privacy-preserving mechanisms, to provide a secure and authenticated communication layer between nodes.
If a function fails during execution, whether due to transient errors, missing state, or platform-level issues, the orchestrator detects the failure and transparently reschedules the function on an eligible node. Since Databelt functions receive state explicitly via immutable Databelt State Keys and avoid side effects, function executions are designed to be idempotent. This ensures that retries do not lead to duplicated computation or inconsistent state. In the case of node loss or network partition, Databelt relies on continuous topology monitoring to update its control-plane graph and exclude unreachable nodes from future state placement decisions. While this mechanism ensures adaptation to dynamic environments, network monitoring, failure detection, and recovery are considered external responsibilities and not part of Databelt’s architecture and mechanisms.


\section{Related Work} \label{sec:relatedw}

In this section, we review related work in three key areas: serverless computing in the 3D Continuum, stateful serverless systems, and orbital task scheduling.

\subsection{Serverless in the 3D Compute Continuum}

Krios~\cite{Krios} introduces a scheduling abstraction to enable targeted application deployment based on satellite mobility on LEO satellites. It uses satellite path prediction to proactively offload applications in specific LEO zones, reducing the number of replicas needed while maintaining availability and minimizing bandwidth and latency overheads. In contrast to Databelt, Krios is not designed for serverless and is limited to LEO-only environments. Moreover, Krios does not incorporate workflow-level state management and SLO-awareness.
Pfandzelter et al.~\cite{LeoComputingPlatform2021} propose that a LEO Edge computing platform has requirements such as low overhead and elastic scalability, and conclude that serverless computing is the best fit.
Nevertheless, there are, to the best of our knowledge, very few serverless platforms designed specifically for LEO.
Komet~\cite{komet2024} is a serverless platform for LEO Edge services, which migrates the data used by a serverless function periodically to the next satellite that is going to be in direct range of the client to minimize latency.
However, Komet only considers single functions and their clients; the latter are assumed to be on Earth.
This is a much smaller scope than Databelt, which considers the entire serverless workflow and optimizes data placement to meet invocation latency SLOs between any two functions of the workflow, both of which may be running on satellites.
Lin et al.~\cite{ResilientAccessEquality6G_LEO2023} propose a 6G serverless Edge platform for running ML applications on satellite swarms and terrestrial nodes.
The platform relies on software-defined networking controllers that integrate ML algorithms for optimization.
However, the authors do not specify how the function state or data is handled.

\subsection{Stateful Serverless}
Stateful serverless frameworks can be classified according to where they provide the statefulness: at the function level or at the workflow level.
Common among all frameworks is the use of some form of key-value store to persist state.

\emph{Function-level statefulness} systems provide mechanisms for functions to directly maintain state.
While this provides added flexibility to developers, it has the disadvantage that the orchestrator typically does not know which state is needed by a function, hence, it cannot proactively load it onto a node.
Additionally, this approach causes a mix of responsibilities for functions because they do not focus on business logic alone but also on state maintenance.
Unlike Databelt, Cloudburst~\cite{Cloudburst2020}, MISO~\cite{goronjic2024miso}, and Crucial~\cite{Crucial2022} provide function-level statefulness.
Cloudburst and MISO allow functions to store elementary data types, arrays, and/or sets using a key-value approach.
Crucial allows the definition and storage of objects as state, which increases productivity for the developers.
All systems, except MISO, store data in high-availability key-value databases.
MISO relies on Conflict Free Replicated Data Types (CRDTs) to improve speed versus conventional key-value databases.
Cloudburst uses the lock-free Anna KVS~\cite{anna} to increase speed.

\emph{Workflow-level statefulness} systems, like Databelt, provide the state management mechanisms in the workflow orchestrator, keeping the serverless functions themselves stateless.
Since the orchestrator knows which state is needed by which function, it can proactively load the required data onto the respective compute nodes.
Workflow-level statefulness also forces a clearer separation of concerns because functions are only responsible for their business logic, while state management is handled by the orchestrator.
AWS Step Functions~\cite{AWS_StepFunctionsOverview, StepFunctionsRedrive2023} and Azure Durable Functions~\cite{DurableFunctions2021, Netherite2022} allow using variables in a workflow to maintain state, which can be passed to functions.
Like Databelt, these systems can store arbitrary state data, but they are designed for the Cloud and do not consider the network topology for their operations.

\subsection{Stateful Serverless at the Edge}
Bringing stateful serverless to the Edge comes with various challenges, including bandwidth restrictions and less storage space per node.
In~\cite{EnergyEffStatefulFaaSEdge2024}, the authors assign each application instance (or session) to a function instance in WebAssembly that directly keeps its state.
For successor functions, they prefer the node that hosts the predecessor to save on network traffic, but they do not specifically address how to optimize state transfer between nodes.
Keeping state local within a function has also been implemented in production, e.g., by Azure Entity Functions~\cite{AzureEntityFunctions} and Cloudflare Durable Objects~\cite{CloudflareDurableObjects} in the Cloud or AWS Long-lived Functions for the Edge~\cite{AWS_GreenGrassLongLivedFunctions}.
Puliafito et al.~\cite{StatefulFaasAtEdge2022} allow functions to switch dynamically between fetching their state from a remote store or keeping their state locally.
Cicconetti et al.~\cite{FaasModelForEdge2022} discuss and evaluate three schemes for managing state on the Edge for serverless function chains, and with some modifications, for DAG-based workflows.
They propose i) pure FaaS, where each function returns the state to the caller of the workflow, who must, then, pass it to the next function, ii) StateProp, where the state is propagated among the functions themselves using arguments and return values, and iii) StateLocal, where the state of a function remains on the node that executed the function and only a pointer is passed to subsequent functions, which can transfer the state to their node if needed.
Their evaluation shows that StateProp reduces communication overhead and end-to-end latency up to a certain workflow size.
For larger workflows, StateLocal, which is actually a local-remote hybrid because it allows migrating state to another node, has proven more beneficial.
Our work builds upon those insights by informing our decision process on handling state in different scenarios. 
MISO~\cite{goronjic2024miso} automatically replicates all data belonging to a function to all nodes that run an instance of this function using CRDT.
Replicating to all nodes is not possible in the 3D~Continuum -- Databelt chooses where to propagate state data based on the workflow and the network topology.
DS2P~\cite{DS2P2024} is a system that allows defining a maximum data access latency SLO for the state of a serverless function.
To fulfill the SLO, D2SP maintains a data distance table on every node that estimates the latency to each relevant key-value pair and only considers nodes for scheduling if they host at least one key-value pair required by a function.
Such a data distance table on every node would require considerable effort to maintain in the highly dynamic 3D~Continuum, hence, it is not an option for Databelt.
Similar to Databelt, D2SP supports migrating data between nodes to minimize SLO violations.
D2SP also seeks to minimize data migration time, which is not relevant when using LEO satellite connections, because they are typically high bandwidth.
MISO and DS2P bear similarities to Databelt, but they lack consideration of when nodes come into and go out of range, which is important for satellites.

\subsection{Serverless Function Fusion}
Fusion of serverless functions is often used to reduce cold start time by avoiding the initialization of a new sandbox. 
Octopus~\cite{wang2025octopus} introduces a fusion node state management system that groups multiple function states at the node level to reduce state I/O operations and enable warm function reuse. However, it still relies on frequent lookup table accesses, which can become a bottleneck under high concurrency, and does not optimize workflow state placement, requiring functions to fetch state from remote services. In contrast, Databelt precomputes and propagates state locations based on the workflow structure, allowing each function to know its state location at invocation time and placing the state locally or nearby, thereby minimizing remote accesses and reducing function state overhead.
SAND~\cite{SAND2018} runs all functions that belong to the same application in a per-application sandbox, instead of a per-function sandbox. 
Additionally, SAND employs a hierarchical message bus, s.t. messages between functions on the same node do not have to traverse the network.
WiseFuse~\cite{WISEFUSE2022} optimizes the DAG of a serverless workflow by fusing consecutive functions based on the benefit of their cumulative latency and cost and by bundling parallel executions of the same function together to allow stragglers to benefit from additional compute resources after faster instances have finished.
However, unlike Databelt, it requires functions to be profiled before being able to optimize them.
FUSIONIZE~\cite{2024fusionize} is a framework that continuously monitors the execution of a serverless workflow and dynamically fuses functions into function groups based on a heuristic.
CWASI~\cite{Cwasi2023} is a container runtime shim for WebAssembly that fuses trusted functions that have been assigned to the same node into the same sandbox to allow direct exchange of data.
While these approaches reduce the number of sandboxes that need to be started and some avoid network communication for co-located functions, each function still needs to read/write its state independently, which results in increased latency if the state needs to be fetched from another node.
Databelt not only fuses the functions into a single sandbox and serves state requests from local storage, if available, but also fuses the functions' state read/write requests, thus reducing the number of network requests.


\section{Conclusion and Future Work} \label{sec:conclusion}

In this paper, we introduced Databelt, a state-aware serverless workflow model and architecture designed to optimize function state propagation in serverless workflows across the Edge-Cloud-Space 3D Continuum. Databelt identifies environmental properties such as the space network topology and proactively moves function states in orbit to the most suitable execution node, thus minimizing latency and inter-node communication hops. Additionally, Databelt introduces a function state fusion mechanism, which reduces redundant storage operations by fusing function states within a shared function sandbox.

Our evaluation shows that Databelt significantly improves serverless workflow execution latency by up to 66\% and increases throughput by 50\% compared to stateless approaches. The function state migration mechanism improves local state availability to 79\%, ensuring that function states are placed locally in the nodes that the target functions execute, ensuring that the functions in the workflow meet predefined SLO constraints. Furthermore, the function state fusion mechanism reduces latency further by up to 20\%, avoiding multiple requests to retrieve the function state. These results demonstrate the benefits of Databelt in facilitating efficient serverless workflows within dynamic environments such as the Edge-Cloud-Space 3D Continuum. Databelt enables serverless workflows to adjust to the limitations and challenges of in-orbit processing by decreasing dependence on remote storage services and consequently enhancing overall workflow performance.

In the future, we plan to expand Databelt by introducing zero-copy mechanisms to optimize the state transfer between the functions. To achieve this, we plan to develop a selection mechanism that dynamically determines the best state transfer approach based on network topology, node proximity, and available bandwidth. 
Moreover, we intend to implement a lightweight, serialization-free state transfer mechanism that prevents unnecessary state persistence by identifying whether a function requires remote state access. 
Finally, we plan to integrate reinforcement learning (RL) into Databelt to optimize AI model deployments and proactively migrate them across the 3D Continuum, ensuring that AI models are positioned near the executing functions. Thus, it minimizes cold start latency, reduces transfer overhead, and enhances the execution time of AI-driven serverless workflows.

\section*{Acknowledgment}
This work is partially funded by the Austrian Research Promotion Agency (FFG) under the project RapidREC (Project No. 903884).
This work has received funding from the Austrian Internet Stiftung under the NetIdee project LEO Trek (ID~7442).
This research received funding from the EU’s Horizon Europe Research and Innovation Program through TEADAL (GA No. 101070186) and NexaSphere projects (GA No. 101192912).

\balance

\bibliographystyle{ACM-Reference-Format}
\bibliography{references}

\end{document}